\documentclass[aps,prb,onecolumn,preprint,tightenlines,superscriptaddress,amsmath,amssymb,eqsecnum]{revtex4-1}

\usepackage{graphicx,color}
\usepackage{amsmath,amssymb}
\usepackage{bm}

\DeclareMathOperator{\sgn}{sgn}

\newcommand{\tr}{\mbox{tr}}

\newcommand{\oprod}[2]{\vert{#1}\rangle\langle{#2}\vert}

\newcommand{\Id}{I}

\newcommand{\bvec}[1]{\bm{#1}}
\newcommand{\bmin}{\bvec{b}_{\mathrm{min}}}
\newcommand{\n}{N}
\newcommand{\gen}{X}
\newcommand{\Gen}{Y}

\newcommand{\coords}{{\tilde{\theta}}}

\newcommand{\var}[1]{\Delta{#1}^2}

\newcommand\QCRB{\hbox{QCRB-O}}

\newcommand{\tw}{{\tt w}}
\newcommand{\minustw}{{\tt-\tw}}
\newcommand{\tz}{{\tt z}}
\newcommand{\minustz}{{\tt-\tz}}
\newcommand{\pmtz}{{\tt\pm\tz}}
\newcommand{\twone}{{\tt w^1}}
\newcommand{\minustwone}{{(\tt-\tt w)^1}}
\newcommand{\tone}{{\tt1}}
\newcommand{\tminusone}{{\tt-\tt1}}
\newcommand{\tzero}{{\tt0}}

\newcommand{\ie}{{\it i.e.}}

\definecolor{green}{rgb}{0.25,0.50,0.25}

\newcommand{\revision}[1]{#1}

\begin{document}

\title{One from many: Estimating a function of many parameters}

\author{Jonathan A. Gross}\email{jarthurgross@gmail.com}
\affiliation{Center for Quantum Information and Control, University of New Mexico, Albuquerque NM 87131-0001, USA}
\affiliation{Centre for Engineered Quantum Systems, School of Mathematics and Physics, The University of Queensland, St.~Lucia QLD 4072, Australia}
\affiliation{Institut quantique and D\'epartment de Physique, Universit\'e de Sherbrooke, Qu\'ebec J1K 2R1, Canada}
\author{Carlton M. Caves}\email{ccaves@unm.edu}
\affiliation{Center for Quantum Information and Control, University of New Mexico, Albuquerque NM 87131-0001, USA}

\date{\today}

\begin{abstract}
Difficult it is to formulate achievable sensitivity bounds for quantum
multiparameter estimation.  Consider a special case, one parameter from
many: many parameters of a process are unknown; estimate a
specific linear combination of these parameters without having the
ability to control any of the parameters.  Superficially similar
to single-parameter estimation, the problem retains genuinely multiparameter
aspects.  Geometric reasoning demonstrates the conditions, necessary and
sufficient, for saturating the fundamental and attainable \revision{quantum-process} 
bound in this context.
\end{abstract}

\maketitle

\section{Introduction}
\label{sec:introduction}

Well-traveled is the path of deriving quantum bounds on the \revision{mean-square error} of estimating a single parameter.
Fisher information\cite{Fisher1922b,Dugue1937a,Rao1945a,Cramer1946a,vanTrees2001a} provides the necessary concept.
Marrying Fisher information to quantum measurement theory---a marriage made in heaven!---yields the quantum Cram{\'e}r-Rao bound (QCRB) on estimating a single parameter.\cite{Helstrom1976a,Wootters1981a,Holevo1982a,ref-braunstein_statistical_1994,Braunstein1996a,ref-boixo_generalized_2007}
Less traveled is the deceptively similar trail of estimating a function of several parameters.
Similar, yes, yet not merely a recasting of the single-parameter problem, this is a different problem with genuinely multiparameter connotations.

For those venturing onto this path, this paper formulates a roadmap for navigating the tricky terrain.
Our work, challenged into existence by Eldredge {\it et al.},\cite{ref-eldredge_optimal_2016} explores
the bound, presented there, on estimating a function of the parameters.  Our goals: examine and interpret
this bound, relating it to the standard bound on estimating a single parameter; formulate
the quantum version in terms of a~QCRB; find the necessary and sufficient conditions for
saturation of the quantum bound; \revision{finally, optimize over quantum measurements and
states to forge a new bound that depends only on the quantum process that imprints the information
one wants to determine.}  The key to achieving these goals comes, surprisingly,
from differential geometry: respect the distinction between tangent vectors, \revision{associated with single-parameter estimation,} and differential
forms, \revision{associated with estimation of a function}, a distinction obscured and suppressed by a parochial preoccupation with single-parameter
estimation.

\revision{Work within the physics community\cite{ref-eldredge_optimal_2016,WGe2018a,qian_heisenberg_2019,paris_quantum_2009,Proctor2018a,Rubio2020a,sekatski_optimal_2020} has considered the task we set for ourselves here.  Distributed (or networked) quantum sensing is how physicists describe this task,
thinking that the function to be estimated is constructed from parameters on distributed sensing devices.
Sidhu and Kok\cite{sidhu_geometric_2019} provide an overview of distributed sensing in Sec.~VIII of an excellent review of quantum parameter estimation.
To avoid the pitfalls of single-parameter thinking, a typical approach is to
calculate genuine multiparameter-estimation bounds and from these to extract
a function-estimation bound.\cite{paris_quantum_2009,Proctor2018a,Rubio2020a}.
Successful though this approach is, it obscures the geometry of the problem
through the introduction of extraneous ingredients and suffers from uncertainty
in the saturability of some bounds.  Both issues we address, by identifying the relevant geometric objects.}

\revision{An extensive statistics literature has considered Fisher-information bounds
on estimating one or more relevant parameters in the presence of a set of irrelevant parameters called nuisance parameters.
Developed for classical estimation in the 1970s through
1990s,\cite{Efron1977a,Godambe1980a,Godambe1984a,Kumon1984a,Amari1987a,Amari1988a,Bhapkar1989a,Bhapkar1994a,YZhu1994a} this nuisance-parameter approach
has been extended recently to quantum estimation.\cite{gazit_quantum_2019,suzuki_nuisance_2019,suzuki_quantum_2020}
Equivalent though the nuisance-parameter language is to what we do here, we generally
avoid it, because it encourages inattention to the distinction between parameter estimation and
function estimation.  Knowing a function to be estimated does not specify a set of irrelevant nuisance
parameters; indeed, such specification defines a single-parameter estimation problem, not a function estimation.
Geometrically invariant language being always our preference, we say, instead of referring to nuisance parameters,
that a subspace of constant function value is not under the control of the experimenter.  Notable also in
this statistics literature, despite formulation as ``information geometry,'' is an absence of geometric
intuition and visualization, as evidenced by the near absence of figures.  That deficiency we remedy.}

\revision{Noteworthy is recent work by Tsang~\textit{et al.},\cite{tsang_quantum_2020} which unites the physics and statistics strands
and considers bounds on function estimation from a geometric perspective.  Similar to, yet different from our analysis,
the work of Tsang~\textit{et al.} is in some ways more general in that the analysis applies to bounds
on quantities in addition to the mean-square error that goes with Fisher information and Cram\'er-Rao bounds.}

\revision{Guided by a single star, this paper rows gently, but steadily and relentlessly in one direction: for an arbitrary,
unitary or nonunitary quantum process, which imprints the information to be estimated on a quantum system, optimize
over quantum measurements and initial state to obtain an achievable QCRB for function estimation that depends only on
the quantum process.  The key geometric object that emerges from this journey we call the process norm.  The process norm
and associated quantum-process bounds on function estimation, along with persistent attention to geometric
thinking and visualization, are the chief contributions of this paper.}

\revision{The general absence of geometric visualization and intuition in the parameter-estimation
literature, invocations of information geometry, is an example of what colleague Christopher
Jackson\cite{csj} calls ``algebra fever,'' the mania in modern mathematics to eschew
geometric thinking in favor of translating geometric concepts into algebra, at which point geometric
intuition is forgotten.}

\revision{Inexplicable, it might be thought, to proceed from Jackson's
observation to an explanation of this paper's distinctive style.  Yet explaining the inexplicable---that's
our job, so listen up.}  Kip Thorne's recent biographical memoir~\cite{Thorne2019}
of John A.~Wheeler reminded us of Wheeler's passion to geometrize Einstein's general relativity
and of the idiosyncratic, yet compelling writing style Wheeler employed to promote that passion.
Possessing a similar passion to geometrize metrology---less grand, to be sure, than Wheeler's
goals, but passionate nonetheless---we adopt here Wheeler's style.  As compelling we hope to be,
but failing that, as idiosyncratic.\cite{Wyler} \revision{The style is a reminder,
on every page of the paper, that we aim to put geometric thinking at the heart of quantum metrology.}

\section{Setting up the problem}
\label{sec:setup}

Specify the problem of interest: estimate a property of a physical process through repeated interactions.
Assume the physical process belongs to a family of quantum channels $\mathcal{E}_{\coords}$ parametrized by $\coords=(\theta^1,\ldots,\theta^\n)$ and the property is a function $q(\coords)$ of these parameters.
Consider interacting with the process by preparing a quantum system in a chosen state, subjecting the system to the evolution the process dictates, and finally measuring the evolved system.
Perform many such interactions, and estimate the property of interest based on the data obtained.
Pose now the natural question: what is the best precision with which the property can be estimated?

A luxury that guarantees proximity to the truth are the many interactions with
the process.  Indulge therefore in an initial estimate for the process encoded
in a parameter point $\coords_0$ near to the true parameter point.  Lock the
interactions to this fiducial operating point.  Precision then describe by
the extent to which small deviations of the truth from $\coords_0$
can be detected.

\revision{Avoid complications arising from singularities and degeneracies by taking the parameters to be independent and physically meaningful in the neighborhood of $\coords_0$.  Such parametrizations are realized by local co\"ordinate charts of the process manifold, as distinct points in this manifold refer to independent, physically distinguishable channels.  More pathological scenarios, induced perhaps by infinite dimensions or additional constraints on the experimenter, might be addressed using techniques employed by Tsang \textit{et al.}\cite{tsang_quantum_2020} for the state-estimation problem; these techniques circumvent direct inversion of potentially singular objects.}

Reify the setup through two examples.  To estimate a phase shift $\varphi$ in the
presence of an unknown loss rate $\gamma$, co\"ordinates $\coords=(\varphi,\gamma)$ parametrize
the process, and the function is simply $q(\varphi,\gamma)=\varphi$.  To estimate
the average fidelity of a process with respect to a target unitary $\mathcal{U}$,
co\"ordinates $\coords$ parametrize the family of all quantum channels, and
$q(\coords)=
\int d\psi\,\big\langle\psi\big\vert\big(\mathcal{U}^{-1}\circ\mathcal{E}_{\coords}\big)
\big(\oprod{\psi}{\psi}\big)\big\vert\psi\big\rangle$.

Unitary processes occupy much of our attention in this exposition, so additional comments
peculiar to this case are in order.  Transform the family of interest $\mathcal{V}_\coords$ to
the equivalent family $\mathcal{U}_\coords=\mathcal{V}_\coords\circ\mathcal{V}_{\coords_0}^{-1}$;
the fiducial parameter point $\coords_0$ then corresponds to the identity
process, $\mathcal{U}_{\coords_0}=\mathcal{I}$.  Convenient it is to parametrize
this unitary by its Hamiltonian $H(\coords)$,
\begin{align}\label{eq:unitaryprocess}
  \mathcal{U}_{\coords}(\rho)
  &=
  e^{-iH(\coords)}\rho e^{iH(\coords)}\,.
\end{align}
The motivating work by Eldredge {\it et al.}\cite{ref-eldredge_optimal_2016} considered
a Hamiltonian for a set of spins and a property $q$, both assumed to be linear in the
chosen parametrization $\coords$,
\begin{align}
  H(\coords)
  &=
  \frac{1}{2}\sum_j\theta^j\sigma^z_j
  =
  \tfrac{1}{2}\theta^j\sigma^z_j\,,
  \\
  q(\coords)
  &=
  \sum_j\alpha_j\theta^j
  =
  \alpha_j\theta^j\,.
\end{align}
The last forms introduce the Einstein summation convention: sum over index
labels that occur simultaneously in a lower and an upper position within an
expression.  Though arbitrary Hamiltonians and properties are not linear
functions of a parametrization, write linear approximations to them in the
neighborhood of the fiducial point $\coords_0$:
\begin{align}\label{eq:Htheta}
  H(\coords_0+d\coords)
  &=
  H(\coords_0)+d\theta^j\left.\partial_jH\right\vert_{\coords_0}
  =
  d\theta^j\gen_j\,,
  \\
  q(\coords_0+d\coords)
  &=
  q(\coords_0)+\left.\partial_jq\right\vert_{\coords_0}d\theta^j
  =
  q(\coords_0)+q_jd\theta^j\,.
\label{eq:qtheta}
\end{align}
Here and throughout, employ the shorthand $\partial_j=\partial/\partial\theta^j$ to
harmonize with the summation convention.  Also eliminated is $H(\coords_0)$, set to
zero since $\mathcal{U}_{\coords_0}=\mathcal{I}$.  Further, reparametrize to choose
$\coords_0=0$ and to set $q(\coords_0)=0$.

Justified indeed are these linear approximations when bounding optimal estimation, as
the limit of many interactions is our concern and the uncertainty in $\coords$
in this limit is correspondingly small.  Press this point home: estimation in
the limit of many experiments is properly studied in the tangent space to the
parameter manifold at a fiducial point.  This perspective we develop in greater
detail in the following section.

\revision{Much attention has been devoted to a problem similar to ours: that of estimating a property of a quantum state with many unknown parameters.\cite{suzuki_nuisance_2019,suzuki_quantum_2020,tsang_quantum_2020}
Only sporadically has the corresponding problem for quantum channels been addressed, a notable example being the estimation of Pauli-channel asymmetry in Gazit {\it et al.}\cite{gazit_quantum_2019}}

\section{Classical estimation: Exercising your differential\\ geometry}
\label{sec:setupc}

Tangent vectors, differential forms, and metrics: these basic elements from
differential geometry provide the mathematical language for the estimation
problem.  Generally forgotten in the parameter-estimation literature are
differential forms, mainly due to a focus on single-parameter problems.  Worthy
of our meditation and attention day and night, renew now acquaintance with these
geometric objects.

\subsection{Classical Fisher information}
\label{sec:Fisher}

Understand first the classical problem of estimating the parameters specifying
a given probability distribution within a parametrized family of distributions
$p(x|\coords)$, reserving for subsequent sections the issue of choosing
initial system state and final system measurement that transform a
parametrized family of quantum channels into such a family of distributions.

The classical procedure is straightforward: sample data~$x$ from the
conditional probability $p(x|\coords)$ and use hatted function $\hat\theta^j(x)$ to
estimate the parameter $\theta^j$ from the data.  The covariance matrix of the
estimators,
\begin{align}\label{eq:Cov}
  C^{jk}
  &=
  \operatorname{Cov}(\hat{\theta}^j,\hat{\theta}^k)
  =
  \int dx\,p(x|\coords)
  \big[\hat\theta^j(x)-\langle\hat\theta^j\rangle_{\coords}\big]
  \big[\hat\theta^k(x)-\langle\hat\theta^k\rangle_{\coords}\big]\,,
\end{align}
captures a \revision{mean-square} notion of the accuracy of the estimates.  In this definition,
\begin{align}
  \langle\hat\theta^j\rangle_{\coords}
  &=
  \int dx\,p(x|\coords)\hat\theta^j(x)
\end{align}
is the mean value of the estimator $\hat\theta^j(x)$, and the parameters
$\coords=(\theta^1,\ldots,\theta^\n)$ should be regarded as true values.

The deviations $\hat\theta^j(x)-\langle\hat\theta^j\rangle_{\coords}$ express
how far the estimates depart from the mean value.  Better it might be thought to
use as deviations the difference between the estimate and the true value,
$\hat\theta^j(x)-\theta^j$; this usage replaces the covariance matrix with the
error-correlation matrix.  An unbiased estimator has mean values equal to true
values, \ie, $\langle\hat\theta^j\rangle_{\coords}=\theta^j$.
Appendix~\ref{app:est-bias} demonstrates how to extract \revision{from a biased estimator an estimator unbiased in a neighborhood of the fiducial operating point, referred to in the literature as a locally unbiased estimator.\cite{suzuki_nuisance_2019,suzuki_quantum_2020,Ragy2016a}}
Specialize now and henceforth to unbiased estimators, thus making the error-correlation matrix identical to the covariance matrix.

The Fisher-information \revision{matrix,\cite{Rao1945a,Cramer1946a,vanTrees2001a}}
\begin{align}\label{eq:fisher-def}
  F_{jk}
  &=
  \int dx\,p(x|\coords)
  \partial_j\ln p(x|\coords)
  \,\partial_k\ln p(x|\coords)
  =
  \int dx\,\frac{1}{p(x|\coords)}
  \partial_jp(x|\coords)
  \,\partial_kp(x|\coords)\,,
\end{align}
is the foundation on which rests classical multiparameter-estimation theory.
For the small deviations from the fiducial operating point contemplated in this paper, the integrands in the expressions for the covariance matrix and the \revision{Fisher-information} matrix should be evaluated at the fiducial point, \ie, $\coords=\coords_0$.

Foundation because the covariance matrix satisfies the matrix inequality
\begin{align}\label{eq:CRB0}
  C\ge F^{-1}\,,
\end{align}
called the multiparameter (classical) Cram{\'e}r-Rao bound (CCRB).\revision{\cite{Rao1945a,Cramer1946a,vanTrees2001a}}
Achieving the CCRB generally requires working in the asymptotic limit of many trials and requires using the right estimators---maximum-likelihood estimation works.
Assume here and hereafter an appropriate estimator and sufficient trials to achieve the \hbox{CCRB}.

\begin{figure}
\centering
\includegraphics[width=0.5\textwidth]{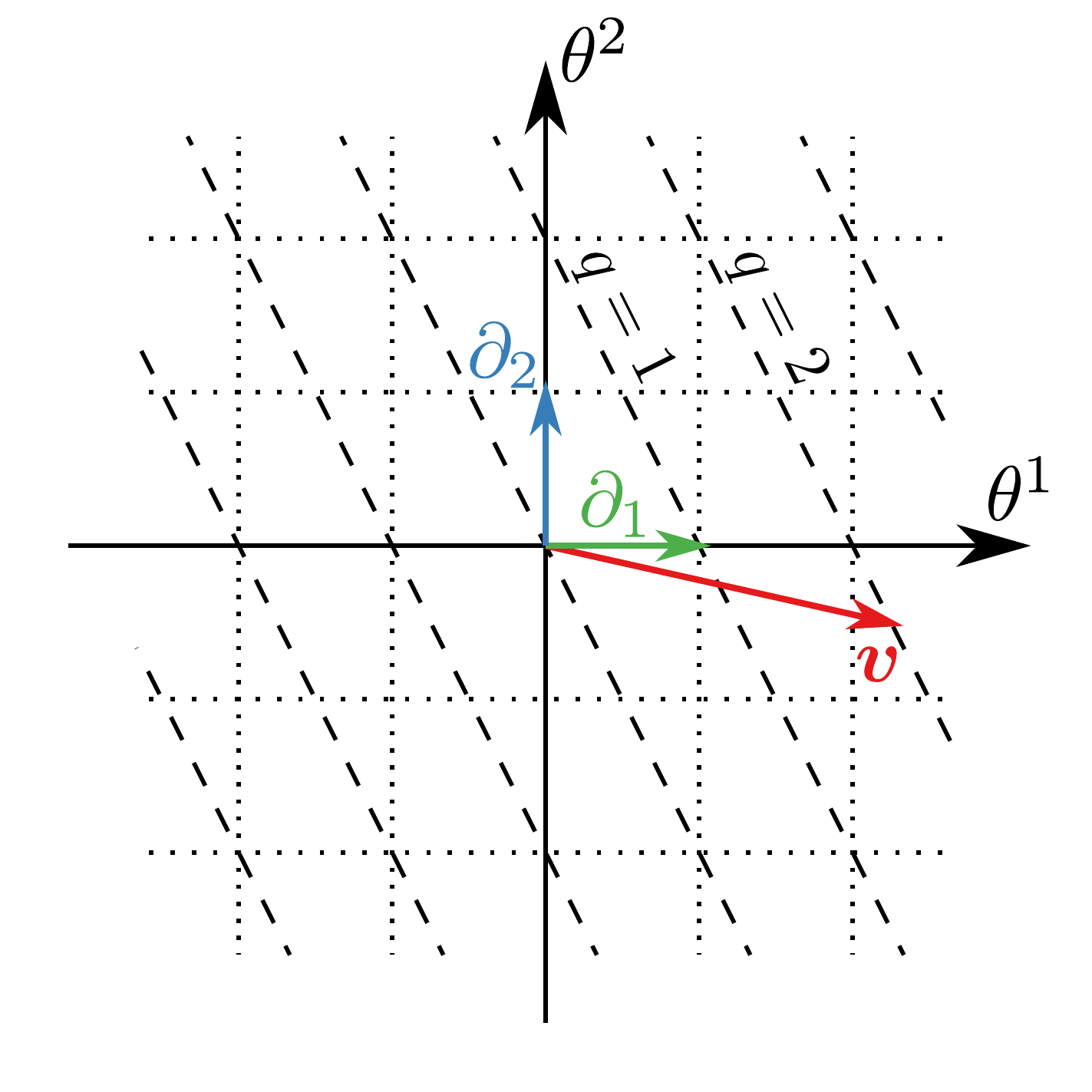}
\caption{Level surfaces of linear combination $q=q_j\theta^j$, here taken to be
$q=\theta^1+\frac12\theta^2$.  This example for parameter~$q$ is used
throughout in figures for two co\"ordinate dimensions (three-dimensional
Fig.~\ref{fig:tangent-match} uses $q=\theta^1+\frac23\theta^2+\frac13\theta^3$).
The vector $\bvec{v}$, here taken to be $\bvec{v}=\frac{9}{4}\partial_1-\frac{1}{2}\partial_2$, extends through two units of $q$, this expressed by $dq(\bvec{v})=\bvec{v}(q)=2$.
The level surfaces of the parameters $\theta^1$ and $\theta^2$ define a square grid.
The directional-derivative basis vector $\partial_{1,2}=\partial/\partial\theta^{1,2}$ lies in a level surface of $\theta^{2,1}$ and extends one unit in $\theta^{1,2}$.
Neatly summarizing this description is the grid equation $d\theta^j(\partial_k)=\partial_k\theta^j={\delta^j}_k$.
No notion of length and orthogonality yet---the square grid for $\theta^1$ and $\theta^2$ is used only for convenience.}
\label{fig:q} \end{figure}

To understand the message of the CCRB, learn now to inhabit the linearized
neighborhood of the fiducial point $\coords_0$.  Of primary importance is
appreciating an important distinction: measuring changes in the property $q$
along a particular path corresponding to varying a linear combination of the
parameters $\theta^j$ requires bringing together two distinct geometric objects,
one that characterizes how $q$ changes as the parameters $\theta^j$ wander
around the neighborhood and another that identifies the particular path.

Call the linearized neighborhood of the fiducial point $\coords_0$ by its formal
name, the tangent space.  Represent a small displacement $\bvec{v}$ on the
tangent space graphically by an arrow, as done in Fig.~\ref{fig:q}, and
algebraically by a directional (partial) derivative,
\begin{align}
  \bvec{v}
  &=
  v^j\partial_j\,.
\end{align}
The vector $\bvec{v}$ is a linear combination of the directional derivatives
$\partial_j$ associated with the co\"ordinates $\theta^j$.

Represent the property $q$ graphically on the tangent space by its level
surfaces, as is done in Fig.~\ref{fig:q}, and algebraically by a differential form,
\begin{align}
  dq
  &=
  q_jd\theta^j\,.
\end{align}
Notice that $dq$ is a linear combination of the differential forms $d\theta^j$
associated with the level surfaces of the co\"ordinates $\theta^j$; together,
the level surfaces of all the co\"ordinates define the familiar co\"ordinate
grid (see Fig.~\ref{fig:q}).

The differential form $dq$ characterizes how $q$ changes in the linear neighborhood
of the fiducial point and is poised to measure the change
in the value of $q$ effected by a vector $\bvec{v}$ in the tangent space:
\begin{align}\label{eq:dqv}
dq(\bvec{v})=\bvec v(q)=v^j\partial_j q=q_jv^j
\end{align}
is the difference between the value of $q$ at the tip of $\bvec v$ and the value
of $q$ at the tail of $\bvec{v}$ (located at the origin).

The parametrization $\coords$ defines a basis of forms and a basis of vectors
dual to one another in the sense that $\partial_j$ lies within the zero surface
of all $d\theta^{k\neq j}$ and extends to the unit surface of $d\theta^j$.
Summarizing the pictorial properties of the co\"ordinate grid is a compact set
of equations:
\begin{align}
d\theta^j(\partial_k)=\frac{\partial\theta^j}{\partial\theta^k}={\delta^j}_k\,.
\end{align}
This formalizes the important distinction: differential forms characterize
how a quantity like $q$ varies in the linear neighborhood of the fiducial point;
vectors specify movement in the tangent space.

Constructed by taking directional derivatives, the Fisher-information matrix~(\ref{eq:fisher-def})
has a natural expression as a (covariant) 2-tensor,
\begin{align}
  \bvec{F}_\downarrow=F_{jk}\,d\theta^j\!\otimes d\theta^k\,,
\end{align}
on the tangent space.  Indeed, manifestly symmetric and positive is the Fisher-information
matrix, so it is a {Riemannian} metric on the tangent space, providing a prescription for
taking inner products between vectors,
\begin{align}
  \langle\bvec{u},\bvec{v}\rangle_F
  &=F_{jk}u^jv^k
  =
  \bvec{F}_\downarrow(\bvec{u},\bvec{v})\,.
\end{align}
The matrix elements $F_{jk}$ are the inner products $\bvec{F}_\downarrow(\partial_j,\partial_k)$.
Positive the Fisher-information matrix \revision{is}, but it can have zero eigenvalues.  Care is required
in dealing with degenerate Fisher-information matrices, as is evident from the CCRB~(\ref{eq:CRB0}).
Proceed now with caution, assuming the Fisher-information matrix is strictly positive;
return to the question of degenerate \revision{Fisher-information} matrices at the end of Sec.~\ref{sec:classical-interpretation}.

\begin{figure}
\centering
\includegraphics[width=0.50000\textwidth]{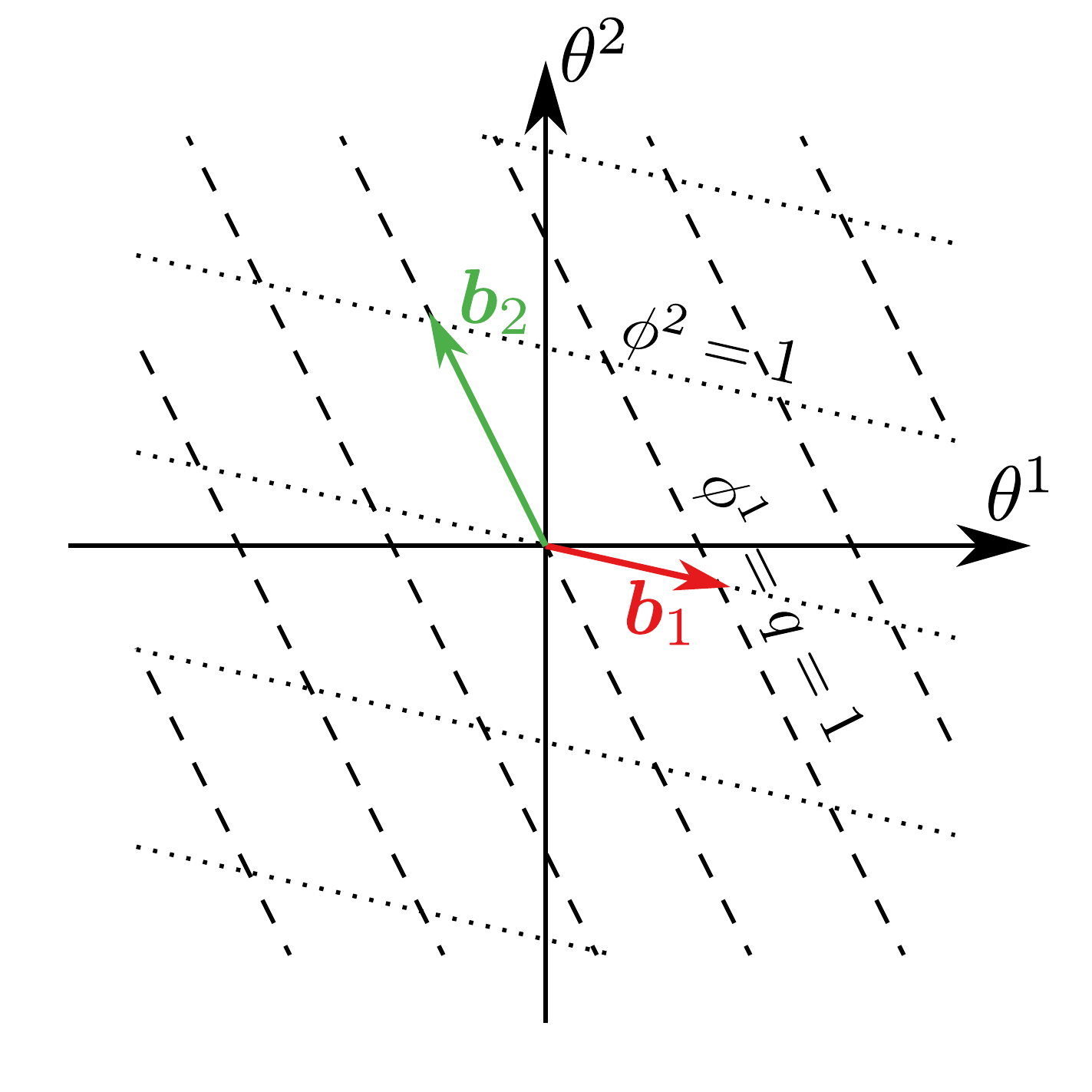}
\caption{Natural co\"ordinate grid relative to the level surfaces of $q$,
natural because one of the parameters, call it $\phi^1$, is equal to $q$.  Read
the $(\phi^1,\phi^2)$ co\"ordinates of a parameter point by identifying the
level surfaces of $\phi^1$ and $\phi^2$ in which the point lies.  The basis
vectors (directional derivatives) are
$\bvec{b}_j=\partial/\partial\phi^j={b^k}_j\partial_k$.  That $\phi^1=q$ means
that $\bvec{b}_1$ advances one unit in $q$ [$dq(\bvec{b}_1)=1$] and that
$\bvec{b}_2$ lies in the $q=0$ level surface means that $dq(\bvec{b}_2)=0$.
Enough to define a single-parameter estimation problem these conditions are not, because
$\bvec{b}_1$, along which $\phi^1$ advances, can point to any location on the
plane $q=1$, its direction determined by the other parameters; specifically, the
direction of $\bvec{b}_1$ is determined by varying $\phi^1$ while holding
$\phi^2$ constant [$d\phi^2(\bvec{b}_1)=0$].
\label{fig:bs}}
\end{figure}

Each vector $\bvec{v}$ defines a single-parameter estimation problem by
locally restricting the family of distributions to parameter variations
that give displacements along $\bvec{v}$.  The Fisher information for
this single-parameter problem is the scalar
\begin{align}\label{eq:Fvv}
\begin{split}
  F_{vv}
  &=
  \int dx\,\frac{1}{p(x|\coords)}
  \bvec{v}\big(p(x|\coords)\big)
  \,\bvec{v}\big(p(x|\coords)\big)
  \\
  &=
  \int dx\,\frac{1}{p(x|\coords)}
  v^j\partial_jp(x|\coords)
  \,v^k\partial_kp(x|\coords)
  =
  \bvec{F}_\downarrow(\bvec{v},\bvec{v})\,.
\end{split}
\end{align}
Pause to savor that the Fisher-information tensor holds within itself the CCRB for
all single-parameter problems.

More explicit we can be about the single-parameter estimation problem specified
by $\bvec v$: vary and estimate a parameter $\phi=\phi^1$ satisfying $d\phi(\bvec{v})=1$,
while holding fixed $N-1$ other parameters $\phi^j$, $j=2,\ldots,N$, satisfying
$d\phi^j(\bvec{v})=0$.  In words, considering $(\phi^1,\ldots,\phi^N)$ as a local
co\"ordinate system, $\bvec{v}$ extends one unit in $\phi$ and points in the
direction obtained by varying $\phi$ while holding the other co\"ordinates
fixed (see Fig.~\ref{fig:bs}); implied is that $\bvec{v}=\partial/\partial\phi$.
The scalar Fisher information~(\ref{eq:Fvv}) bounds the single-parameter estimator
variance (keep in mind the assumption of unbiased estimators,
$\langle\hat\phi\rangle_{\tilde\theta}=\phi$),
\begin{align}\label{eq:var-hatphi}
\var{\hat\phi}
=\int dx\,p(x|\coords)\big[\hat\phi(x)-\phi\big]^2
\ge\frac{1}{F_{vv}}
=\frac{1}{\bvec{F}_\downarrow(\bvec{v},\bvec{v})}\,.
\end{align}
Find a fuller understanding of co\"ordinate systems matched to single-parameter estimation
in Sec.~\ref{sec:scalar-not-single-param}.

The inverse of the Fisher-information matrix is the optimal covariance matrix,
which measures deviations of parameter estimates from the true parameter value.
Since measuring deviations is the job of differential forms, learn with satisfaction
that the natural formulation of the inverse Fisher-information matrix is as a
(contravariant) 2-tensor,
\begin{align}
  \bvec{F}^\uparrow
  =
  (F^{-1})^{jk}\,\partial_j\!\otimes\partial_k
  =
  F^{jk}\,\partial_j\!\otimes\partial_k,
\end{align}
which provides a prescription for
calculating (optimal) covariances of parameters specified by forms,
\begin{align}
  \operatorname{Cov}(\hat{\theta}^j,\hat{\theta}^k)
   =
  F^{jk}
  =
  \bvec{F}^\uparrow(d\theta^j,d\theta^k)
  \,.
\end{align}

\subsection{Scalar estimation}
\label{sec:class-scalar-est}

No control of any of the parameters, no prior constraints on how any
parameter varies, no ability to hold any combination of the parameters
fixed---these mean that an estimate of $q$ must be extracted from estimating
all the \revision{parameters.\cite{paris_quantum_2009,Proctor2018a,Rubio2020a,tsang_quantum_2020}}
Uncertainties in the estimates of all the parameters feed into the uncertainty in the estimate of~$q$.

From the parameter estimators $\hat\theta^j(x)$ comes an estimator $\hat q(x)$,
the same linear combination as $q$ is a linear combination of the parameters
$\theta^j$:
\begin{align}
\hat q(x)=q_j\hat\theta^j(x)\,.
\end{align}
The estimator variance
\begin{align}\label{eq:deltahatq}
\var{\hat q}=\int dx\,p(x|\coords)\big[\hat q(x)-q\big]^2
=q_jq_k\int dx\,p(x|\coords)\big[\hat\theta^j(x)-\theta^j]\big[\hat\theta^k(x)-\theta^k]
\end{align}
---recall the assumption of unbiased estimators, for which $\langle\hat q\rangle_{\coords}=q$---is
the action of the covariance matrix~(\ref{eq:Cov}), written as a contravariant 2-tensor
$\bvec{C}^\uparrow$, on the form $dq$:
\begin{align}\label{eq:Covq}
\var{\hat q}=C^{jk}q_jq_k=\bvec{C}^\uparrow(dq,dq)\,.
\end{align}
The matrix CCRB~(\ref{eq:CRB0}) provides the one-from-many, no-control CCRB
\revision{for the function~$q$},\cite{Efron1977a,Godambe1980a,Godambe1984a,Kumon1984a,Amari1987a,Amari1988a,Bhapkar1989a,Bhapkar1994a,YZhu1994a,gazit_quantum_2019,suzuki_nuisance_2019,suzuki_quantum_2020}
\begin{align}
\var{\hat q}=\bvec{C}^\uparrow(dq,dq)\ge\bvec{F}^\uparrow(dq,dq)\,.
\end{align}
Implicated here is the invariant constructed from the contravariant form of the Fisher metric,
$\bvec{F}^\uparrow$, and the 1-form~$dq$:
\begin{align}\label{eq:Fdqdq}
\bvec{F}^\uparrow(dq,dq)=F^{jk}q_jq_k=q_jq^j\,.
\end{align}
Raise the index on $dq$ using $\bvec{F}^\uparrow$, and find the vector $\bvec{q}_F=q^j\partial_j$ introduced in the last form,
\begin{align}\label{eq:qvec-def}
q^j=F^{jk}q_k\,.
\end{align}
Orthogonal to the level surfaces of $q$, according to the Fisher metric, is $\bvec{q}_F$:
\begin{align}
\langle\bvec{q}_F,\bvec{v}\rangle_F=\bvec{F}_\downarrow(\bvec{q}_F,\bvec{v})=F_{jk}q^jv^k=q_kv^k=dq(\bvec{v})=0\,,
\end{align}
for any $\bvec{v}$ that lies in the level surfaces of $q$.  Express the invariant~(\ref{eq:Fdqdq}) in all its forms,
\begin{align}
\begin{split}
F^{jk}&q_jq_k=\bvec{F}^\uparrow(dq,dq)\\
&=F_{jk}q^jq^k=\bvec{F}_\downarrow(\bvec{q}_F,\bvec{q}_F)=\langle\bvec{q}_F,\bvec{q}_F\rangle_F\\
&=q_jq^j=dq(\bvec{q}_F)\,.
\end{split}
\end{align}
Pause to appreciate that no-control estimation is controlled by this invariant.

\subsection{Scalar estimation is not single-parameter estimation}
\label{sec:scalar-not-single-param}

\begin{figure}
\centering
\includegraphics[width=1.0000\textwidth]{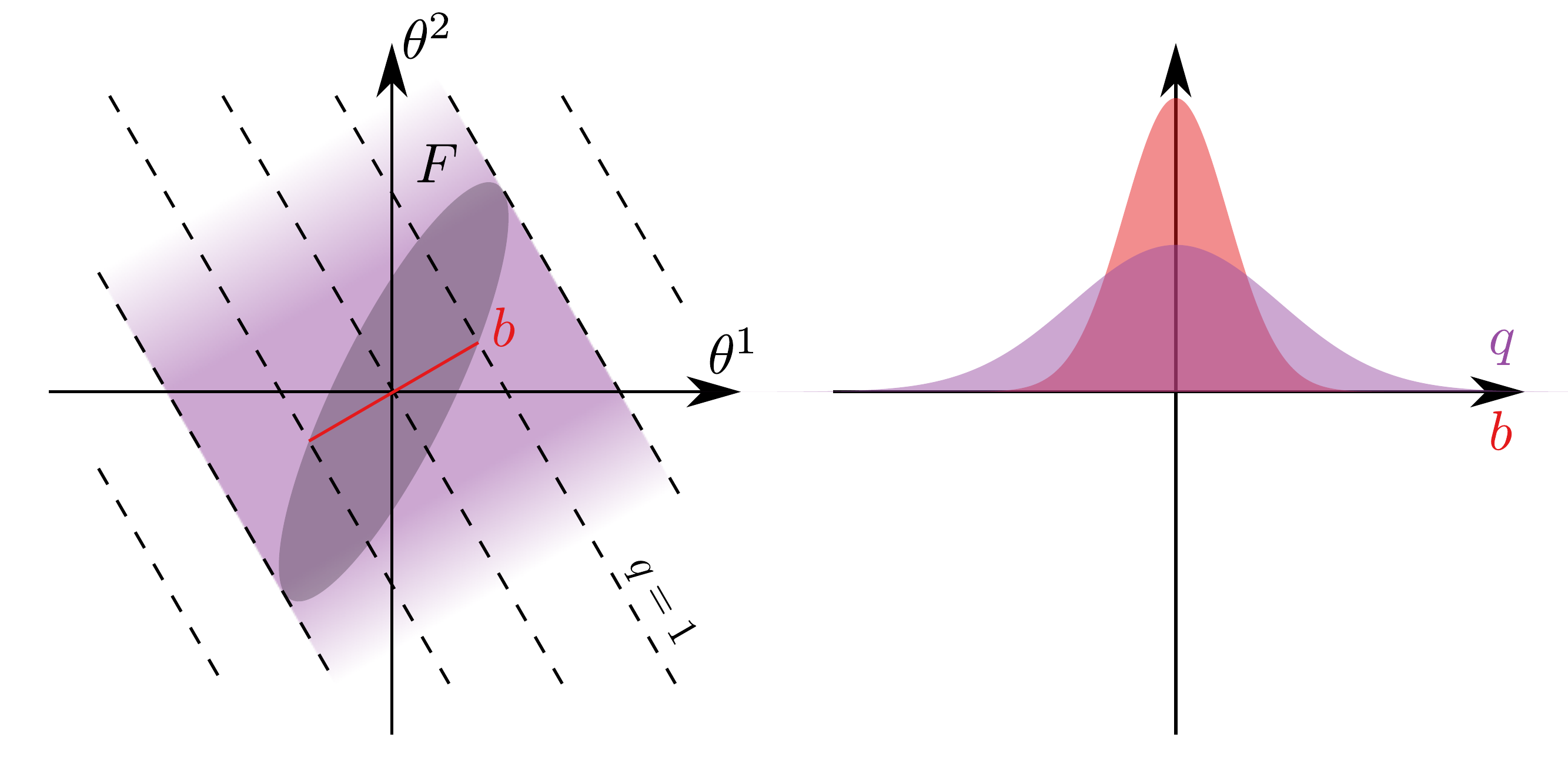}
\caption{The Fisher information for a multiparameter family of probability
  distributions gives the smallest covariance of estimates about the true
  parameter values.  The Fisher information for a single parameter in this
  setting gives the smallest variance of estimates of that parameter about its
  true value, conditioned on knowledge of the true values of the other
  parameters (here illustrated by the parameter $\bvec{b}$).  The smallest
  variance of estimates of the value of a function of the parameters about its
  valuation at the true parameter values is given by the variance of the
  multiparameter-estimate distribution marginalized over parameters that don't
  change the function value (here illustrated by the level surfaces of the
  function $q$).  The probability densities at right illustrate how the
  conditional distribution of a parameter associated with the function values is
  generally narrower than the limit given by the marginal of the Fisher
  information.
\label{fig:marg-cond}}
\end{figure}

Address now the pitfalls in neglecting the distinction between forms and vectors.
Reparametrize the tangent space with new co\"ordinates
$\tilde\phi=\big(\phi^1,\ldots,\phi^\n\big)$.  Match these co\"ordinates to
the job of estimating $q$ by calling out one of the new co\"ordinates,
make it the first, to be $q$ itself, \ie, $\phi^1=q$, with associated
differential form
\begin{align}
  d\phi^1
  &=
  dq
  =
  q_j d\theta^j\,.
\end{align}
Emerging from these new co\"ordinates are new directional derivatives,
\begin{align}
  \frac{\partial}{\partial\phi^j}
  &=
  \bvec{b}_j
  =
  {b^k}_j\partial_k\,,
\end{align}
their vectorial character highlighted by the special designation $\bvec{b}_j$.
Choose \revision{often in the following} to omit the subscript on the special co\"ordinate $\phi^1$ and its
associated directional derivative $\bvec{b}_1$, writing $\phi^1=\phi$ and
$\bvec{b}_1=\bvec{b}$.  These new co\"ordinates and their basis vectors define
a new co\"ordinate grid, characterized by the equations
\begin{align}
  d\phi^j(\bvec{b}_k)=\bvec{b}_k(\phi^j)={\delta^j}_k\,.
\end{align}
One such new co\"ordinate grid is illustrated in Fig.~\ref{fig:bs}.

Suggested by this parametrization is a single-parameter estimation problem closely
tied to the problem of estimating $q$: $\bvec{b}$ specifies a line through
the fiducial origin in the tangent space that specifies a single-parameter
manifold of distributions, and $dq(\bvec{b})=1$ means that the parameter $\phi$
changes by one unit from tail to tip of $\bvec b$.  Alluring though this
identification is, at our disposal are the tools to silence the siren's call.

Observe the difference between optimal variances of single-parameter estimation
of $\phi$ and estimation of $q$ within a multiparameter manifold:
\begin{align}\label{eq:phi1-var-bound}
  \var{\hat{\phi}}
  &\geq
  \frac{1}{\bvec{F}_\downarrow(\bvec{b},\bvec{b})}
  =
  (F_{jk}b^jb^k)^{-1}
  \\
  \var{\hat{q}}
  &\geq
  \bvec{F}^\uparrow(dq,dq)
  =
  (F^{-1})^{jk}q_jq_k\,.
\label{eq:q-var-bound}
\end{align}
Figure~\ref{fig:marg-cond} illustrates the distinction between these two
quantities, depicting the covariance of the full estimator as a shaded ellipse
containing the tips of all vectors $\bvec{v}$ that represent parameter changes
within a standard deviation of the origin, \ie, $\bvec{F}_\downarrow(\bvec{v},\bvec{v})\leq1$.
Variation in $\hat{q}$ is clearly variation in the full estimator distribution
marginalized over deviations that leave $q$ unchanged, while variation in
$\hat{\phi}^1$ is variation in the full estimator conditioned on the other
parameters being held fixed to their fiducial values.

Most importantly, variation in $\hat{\phi}^1$ depends on an arbitrary choice of
parametrization.  Given only the choice $\phi^1=q$, $\bvec{b}_1$ can place its
tip at any point on the plane $q=1$; its direction, required to specify a
single-parameter problem, is determined by the co\"ordinates that accompany
$\phi^1$.  Specifically, $\bvec{b}_1$ points in the direction determined by
holding the other co\"ordinates fixed:
\begin{align}\label{eq:dphijb1}
d\phi^j(\bvec{b}_1)=0\,,\quad j=2,\ldots,\n.
\end{align}
Free we are to modify $\bvec{b}_1$ by adding to it any vector lying in the null
surface of $q$---that is, any linear combination of
$\bvec{b}_2,\ldots,\bvec{b}_\n$. Such a modification of $\bvec{b}_1$ drags along
the co\"ordinates $\phi^2,\ldots,\phi^\n$, ensuring they still satisfy
Eq.~(\ref{eq:dphijb1}).  Different choices for $\bvec{b}_1$ pick out different
single-parameter submanifolds.  The variance of $\hat{\phi}^1$ measures
estimator precision for these irrelevant single-parameter problems. Variation in
$\hat{q}$ rises above petty differences in parametrizations and measures
estimator precision for the no-control problem at hand.

An alternative perspective is that $dq$ and $\bvec{F}^\uparrow$ together
privilege a particular single-parameter problem whose sensitivity bound
coincides with the bound for the scalar estimation problem.  The vector
$\bvec{q}_F$ defined in Eq.~(\ref{eq:qvec-def}), orthogonal to surfaces of
constant $q$ according to the Fisher metric, is not suitably normalized to
define a single-parameter estimation problem, because
$dq(\bvec{q}_F)=q_jq^j=\langle\bvec{q}_F,\bvec{q}_F\rangle_F$.  Suitable it
becomes by scaling it to place the tip on the unit surface of $q$:
\begin{align}\label{eq:bF}
\bvec{b}_F=\frac{\bvec{q}_F}{\langle\bvec{q}_F,\bvec{q}_F\rangle_F}\,.
\end{align}
The vector $\bvec{b}_F$ has squared Fisher length
\begin{align}
\bvec{F}_\downarrow(\bvec{b}_F,\bvec{b}_F)=\langle\bvec{b}_F,\bvec{b}_F\rangle_F=\frac{1}{\langle\bvec{q}_F,\bvec{q}_F\rangle_F}
=\frac{1}{\bvec{F}^\uparrow(dq,dq)}\,,
\end{align}
leading to a no-control CCRB,
\begin{align}
\var{\hat q}\ge\bvec{F}^\uparrow(dq,dq)=\frac{1}{\bvec{F}_\downarrow(\bvec{b}_F,\bvec{b}_F)}\,,
\end{align}
which coincides with the CCRB for single-parameter estimation
defined by $\bvec{b}_F$.

\begin{figure}
\centering
\includegraphics[width=0.50000\textwidth]{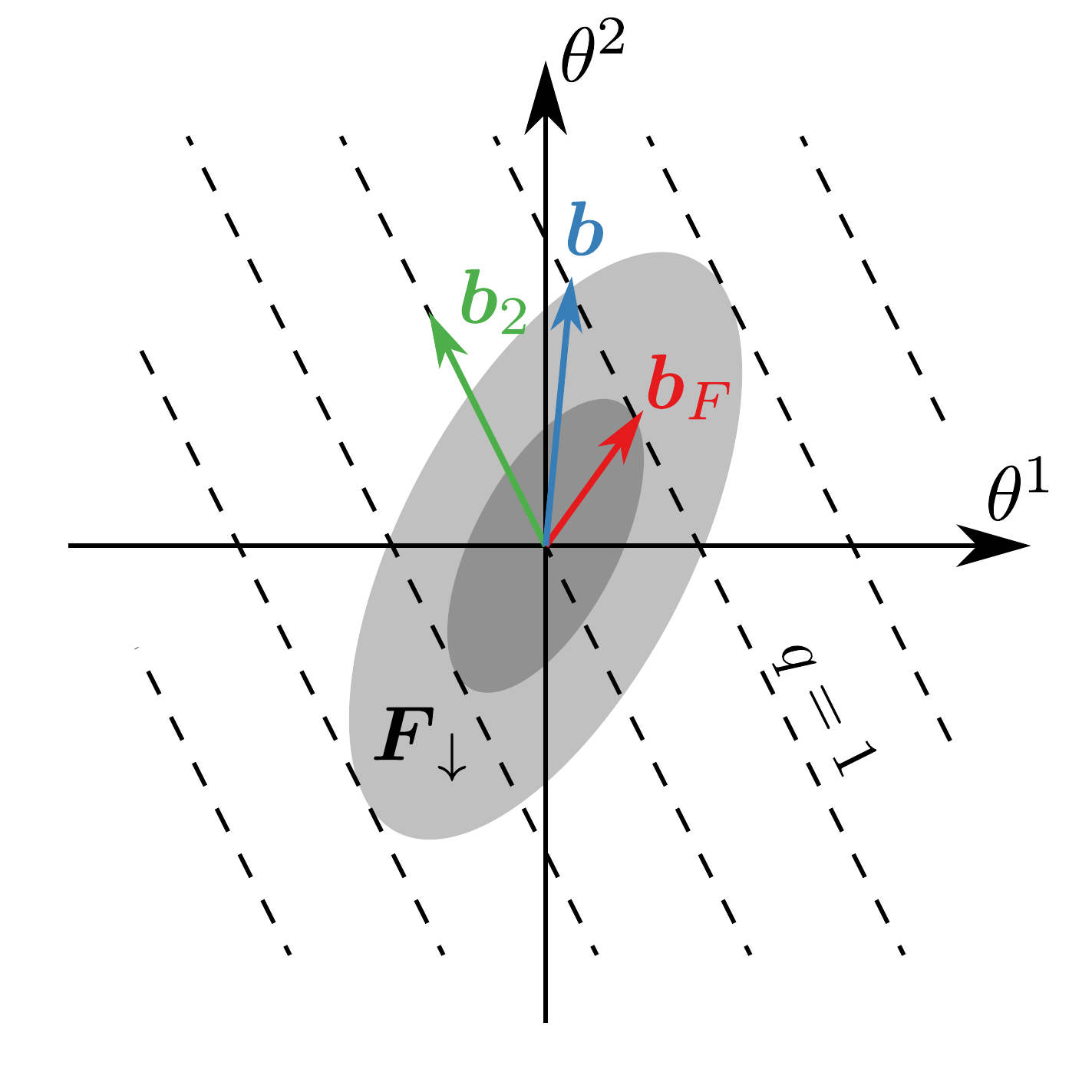}
\caption{Vectors of the same length according to the Fisher metric
$\bvec{F}_\downarrow$ have their tips on a covariance ellipse centered
at the origin.  A single-parameter problem is specified by a vector $\bvec{b}$
that extends one unit in $q$; the Fisher information $F_{bb}$ for this
problem is the length of $\bvec{b}$ as measured by the Fisher metric.
The shortest vector, $\bvec{b}_F$, thus having the least Fisher information,
is orthogonal to the level surfaces of $q$ according to the Fisher metric;
this smallest Fisher information governs estimation of $q$ when one
has no control over any of the parameters $\theta^j$.  Other vectors
that extend one unit in $q$, exemplified by $\bvec{b}$,
have more Fisher information, as they can be made shorter by sliding the
tip along the unit surface of $q$ toward $\bvec{b}_F$.  Indeed, a
way of characterizing orthogonality to the level surfaces of $q$ is that
the tip of $\bvec{b}_F$ is at the point where the Fisher ellipse is
tangent to the surface $q=1$, so that any sliding of the tip of $\bvec{b}_F$
increases the length.\label{fig:cs}}
\end{figure}

Figure~\ref{fig:cs} depicts the geometry: $\bvec{b}_F$, as
the vector orthogonal to level surfaces of $q$ according to the
classical Fisher metric, is the shortest vector that extends
one unit in $q$ and so has the least Fisher information of all such
vectors.  Consider any vector $\bvec{b}$ satisfying
\begin{align}
1=dq(\bvec{b})=q_jb^j=F_{jk}q^jb^k=\langle\bvec{q}_F,\bvec{b}\rangle_F\,.
\end{align}
Cauchy-Schwarz commands,\cite{ref-eldredge_optimal_2016}
\begin{align}\label{eq:CS}
1=\langle\bvec{q}_F,\bvec{b}\rangle_F^2
\le\langle\bvec{q}_F,\bvec{q}_F\rangle_F\,\langle\bvec{b},\bvec{b}\rangle_F
=\frac{\langle\bvec{b},\bvec{b}\rangle_F}{\bvec{F}_\downarrow(\bvec{b}_F,\bvec{b}_F)}\,,
\end{align}
so the Fisher information for $\bvec{b}\ne\bvec{b}_F$ exceeds that for $\bvec{b}_F$,
\begin{align}\label{eq:one-from-many-bound1}
F_{bb}=\langle\bvec{b},\bvec{b}\rangle_F\ge\bvec{F}_\downarrow(\bvec{b}_F,\bvec{b}_F)\,.
\end{align}
Revealed is that the no-control bound is the most pessimistic single-parameter bound:
\begin{align}\label{eq:one-from-many-bound2}
\var{\hat q}\ge\bvec{F}^\uparrow(dq,dq)=\frac{1}{\bvec{F}_\downarrow(\bvec{b}_F,\bvec{b}_F)}\ge\frac{1}{F_{bb}}\,.
\end{align}

\revision{Selection of $\bvec{b}_F$ according to Eq.~(\ref{eq:bF}) is known as ``parameter orthogonalization'' in the statistics literature.\cite{Kumon1984a,Amari1987a,Bhapkar1994a,YZhu1994a,suzuki_nuisance_2019,suzuki_quantum_2020}. A co\"ordinate change, as in the discussion surrounding Eq.~(\ref{eq:dphijb1}), makes $\partial/\partial\phi=\bvec{b}=\bvec{b}_F$ orthogonal, relative to the Fisher metric, to the surfaces of constant $q$.  The co\"ordinate transformation changes only the co\"ordinates other than $q$, so this can be regarded as identifying the right nuisance parameters relative to the Fisher metric, or it can be regarded as finding the single-parameter estimation problem that coincides with no-control estimation of the function~$q$.  Either way, the Cauchy-Schwarz inequality~(\ref{eq:CS}) embodies Fisher orthogonality and thus is the key to selecting $\bvec{b}_F$ as the vector that goes with no-control (function) estimation.}

\subsection{Interpretation}
\label{sec:classical-interpretation}

Apparently identical, yet subtly different, the variances $\var{\hat\phi}$ of Eq.~(\ref{eq:phi1-var-bound}) and $\var{\hat q}$ of Eq.~(\ref{eq:q-var-bound}) teach a lesson: in the integrals~(\ref{eq:var-hatphi}) and~(\ref{eq:deltahatq}) for the variances, the parameters are evaluated at the fiducial point, taken here to be zero parameter values; the difference lies in that $\var{\hat q}$ is honest about its uncertainty in all parameters, whereas $\var{\hat\phi}$ presumes to know the true values of $\phi^2,\ldots,\phi^N$.  Assuming $\hat{\phi}$ has the blind luck to correctly guess $\phi^2,\ldots,\phi^N$, it will outperform $\hat{q}$.  In the presence of real uncertainty, though, $\hat{\phi}$ trips on the tangled web it wove and underperforms $\hat{q}$.

More enlightening still is it to understand, as is depicted in Fig.~\ref{fig:cs}, that the Fisher ellipse $\bvec{F}_\downarrow(\bvec{v},\bvec{v})=\bvec{F}_\downarrow(\bvec{b}_F,\bvec{b}_F)$
is tangent to the unit level surface of $q$.  Implied is that errors in estimates of parameters
that don't change $q$ are uncorrelated with errors in $\bvec{b}_F$; there is no danger in using an
estimator that assumes incorrect values for such parameters.  This insensitivity to errors in
the other parameters is the reason the single-parameter problem specified by $\bvec{b}_F$ is
the same as the no-control estimation problem for $q$: a single-parameter problem assumes the
other parameters are fixed at their fiducial values, but for the special single-parameter
problem specified by $\bvec{b}_F$, this assumption is unnecessary, and the other parameters
can be left uncontrolled.

Insensitivity to errors in these other parameters suggests considering
\revision{Fisher-information} matrices that are degenerate and thus not metrics at
all.  Of particular interest is a rank-one \revision{Fisher-information} matrix,
\begin{align}\label{eq:Fdegenerate}
\bvec{F}_\downarrow=A^2\,dq\otimes dq\,,
\end{align}
where $A$ is a constant.  The components of the \revision{Fisher-information} matrix are
\begin{align}
F_{jk}=\bvec{F}_\downarrow(\partial_j,\partial_k)
=A^2\,dq(\partial_j)\,dq(\partial_k)
=A^2q_jq_k\,.
\end{align}
Constructed from $dq$ alone, the \revision{Fisher-information} matrix~(\ref{eq:Fdegenerate}) enjoys
the exalted status of the invariant
$\bvec F^\uparrow(dq,dq)=1/\bvec F_\downarrow(\bvec b_F,\bvec b_F)$.
Any vector $\bvec v$ has Fisher information
\begin{align}\label{eq:Fdegenerate2}
\bvec{F}_\downarrow(\bvec{v},\bvec{v})=F_{vv}=A^2(q_jv^j)^2=A^2[dq(\bvec v)]^2\,,
\end{align}
meaning that any sampling procedure giving rise to such a \revision{Fisher-information} matrix is
sensitive only to the parameter $q$ and not to any of the other co\"ordinates.
Indeed, any vector $\bvec{b}$ satisfying $dq(\bvec{b})=1$ has Fisher information
\begin{align}\label{eq:Fdegenerate3}
\bvec{F}_\downarrow(\bvec{b},\bvec{b})=F_{bb}=A^2\,,
\end{align}
making this \revision{Fisher-information} matrix the embodiment of one-from-many estimation: no matter
what are the co\"ordinates other than $\phi=q$, the Fisher information is the
same (all $\bvec b$ have the same Fisher length).

The Fisher ellipsoid degenerates to a pair of level surfaces of $q$ having
opposite values of~$q$.  Equivalent to Eq.~(\ref{eq:Fdegenerate3}) is that for any
vector $\bvec v$ that lies in the level surface $q=0$, \ie, $dq(\bvec{v})=0$,
\begin{align}
\bvec{F}_\downarrow(\bvec{v},\bvec{v})=0=\bvec{F}_\downarrow(\bvec{v},\bvec{b})\;.
\end{align}
Quantum procedures that yield these \revision{Fisher-information} matrices come up in Sec.~\ref{sec:commuting-generators}.

\section{Quantum estimation}
\label{sec:setupq2}

\subsection{Quantum Cram{\'e}r-Rao bound}
\label{sec:QCRB}

Return now to the quantum setting, abandoned at the end of Sec.~\ref{sec:setup}.  Quantum mechanics generates
the classical conditional probability $p(x|\coords)$ from an initial state $\rho$, which is processed
through a quantum process $\mathcal{E}_\coords$ to give a state,
\begin{align}\label{eq:rhotheta}
\rho_\coords=\mathcal{E}_\coords(\rho)\,,
\end{align}
and a measurement described by a POVM $\{E_x\}$, whose outcome $x$ is the data collected by the measurement:
\begin{align}
p(x|\coords)=\tr(E_x\rho_\coords)\,.
\end{align}
Appreciate that in the quantum setting, the Fisher-information matrix and its Fisher ellipsoid are
functions of the initial state $\rho$ and the quantum measurement used to extract data from the system.

The foundation of quantum \revision{estimation of a single parameter} is the quantum
Fisher \revision{information,\cite{Helstrom1976a,Holevo1982a,Wootters1981a,ref-braunstein_statistical_1994}} defined at the fiducial state $\rho=\rho_0$ by
\begin{align}\label{eq:Qbb}
  Q_{bb}&=\max_{\text{measurements}|\rho_0}F_{bb}=\tr(\rho_0L_b^2)\,, \\
  \bvec{b}\rho_\coords\big|_{\coords=0}&=\left.\frac{\partial\rho_\coords}{\partial\phi}\right|_{\coords=0}
  =\frac{1}{2}(\rho_0L_b+L_b\rho_0)\,.
  \label{eq:SLD}
\end{align}
The Hermitian operator $L_b$ sports the title of symmetric logarithmic derivative (SLD);
notice that $\tr(\rho_0 L_b)=0$.  Foundation the quantum Fisher information is because
according to Eq~(\ref{eq:Qbb}), it is the same as the classical Fisher
information for the best quantum measurement;\cite{ref-braunstein_statistical_1994}
hence, find the bound
\begin{align}\label{eq:single-param-quant-bound}
F_{bb}\le Q_{bb}\,.
\end{align}
The result is a chain of bounds on estimator variance:
\begin{align}\label{eq:singlebound2}
\var{\hat\phi}\ge\frac{1}{F_{bb}}\ge\frac{1}{Q_{bb}}\,.
\end{align}
The chain can be saturated: the first inequality, asymptotically in many trials,
by using, for example, maximum-likelihood estimation; the second by choice of
optimal quantum measurement.

Consider now unitary operations, as in Eq.~(\ref{eq:unitaryprocess}), where
$\rho_\coords=\mathcal{U}_\coords(\rho)$.  The \hbox{Hamiltonian}~(\ref{eq:Htheta}),
written in terms of the new parameters~$\tilde\phi$
and associated generators, becomes
\begin{align}\label{eq:Hphi}
H(\coords)=\phi^j\Gen_j=\phi\Gen+\sum_{j=2}^\n\phi^j\Gen_j=H(\tilde\phi)\,.
\end{align}
Here
$\Gen_j=\partial H/\partial\phi^j=(\partial\theta^k/\partial\phi^j)X_k=
{b^k}_jX_k$.  Generating changes in $\phi=q$ is the operator
\begin{align}
\Gen=\Gen_1=b^k\gen_k=\bvec{b}H(\coords)
=
\frac{\partial H(\coords)}{\partial\phi}\,.
\end{align}

Cumbersome indeed is the implicit expression~(\ref{eq:SLD}) for determining the
SLD $L_b$, but an appealingly simple, explicit form is available for a unitary
process,
$\rho_\coords=\mathcal{U}_\coords(\rho)=e^{-iH(\coords)}\rho\,e^{iH(\coords)}$,
applied to a pure fiducial state $\rho=|\psi\rangle\langle\psi|$.  For a unitary
process, it is always true that
\begin{align}
\left.\frac{\partial\rho_\coords}{\partial\phi}\right|_{\coords=0}
=-i[\Gen,\rho]=-i[\Delta\Gen,\rho]\,;
\end{align}
introduction of the operator deviation
$\Delta\Gen=\Gen-\langle\Gen\rangle=\Gen-\tr(\rho\Gen)$ makes life easier
shortly.

Realize now that for a pure fiducial state, since
$\rho_\coords=\rho^2_\coords$, have we
\begin{align}
\frac{\partial\rho_\coords}{\partial\phi}
=\rho_\coords\frac{\partial\rho_\coords}{\partial\phi}+\frac{\partial\rho_\coords}{\partial\phi}\rho_\coords
\end{align}
and the conclusion,
\begin{align}
L_b=2\left.\frac{\partial\rho_\coords}{\partial\phi}\right|_{\coords=0}
=-2i[\Delta\Gen,\rho]\,.
\end{align}
\revision{(Note that the SLDs are not completely determined for rank-deficient states, since their projection onto the null space of $\rho$ is irrelevant.\cite{fujiwara_quantum_2016})} Simple now is the quantum Fisher information~(\ref{eq:Qbb}):
\begin{align}
Q_{bb}=-4\,\tr\big((\Delta\Gen\rho-\rho\Delta\Gen)^2\rho\big)
=4\langle(\Delta \Gen)^2\rangle\,.
\end{align}
The variance of the generator $\Gen$ is calculated in the fiducial
(pure) system state $\rho$.

Confronting us again, now in the quantum setting, are the requirements for
defining a single-parameter estimation problem.  The
generator $\Gen=\bvec bH(\coords)$, whose variance is the quantum
Fisher information, is determined by the vector $\bvec{b}$ that defines
the single-parameter problem.  The Hamiltonian~(\ref{eq:Hphi}) emphasizes
that the parameters that accompany $\phi$ must be held fixed to get a clean
estimate of $\phi=q$.

One more inequality,
\begin{align}\label{eq:seminorm}
Q_{bb}=4\langle(\Delta \Gen)^2\rangle\le\Vert\Gen\Vert_s^2=\Vert\bvec b
H(\coords)\Vert_s^2\,.
\end{align}
completes the quantum discussion, by introducing the operator
seminorm~$\Vert\Gen\Vert_s$,\cite{ref-boixo_generalized_2007}
the difference between the largest and smallest eigenvalues of $\Gen$.
Add yet one more bound to the chain of single-parameter estimator
bounds~(\ref{eq:single-param-quant-bound}),
\begin{align}\label{eq:singlebound3}
\var{\hat\phi}\ge\frac{1}{F_{bb}}\ge\frac{1}{Q_{bb}}\ge\frac{1}{\Vert\bvec{b}H(\coords)\Vert_s^2}\,.
\end{align}
Saturated is the last inequality by choosing an optimal fiducial state, an equal
superposition of the eigenstates of $\Gen$ with largest and smallest eigenvalues.
Equally deserving the appellation of quantum Cram{\'e}r-Rao bound (QCRB) are the last
two inequalities; distinguish them by letting the first be the QCRB and the
second, the focus of our attention because of its optimal fiducial-state, the \QCRB.

Quantum Fisher information also comes in a multiparameter version, in which it
is a positive matrix that defines a quadratic form on the space of parameters.
With no need for this quantum \revision{Fisher-information} matrix, tarry not to introduce it.  The quantum
\revision{Fisher-information} matrix enjoys only a vestigial presence in our treatment: labeling the
single-parameter quantum Fisher information as the $bb$ component of a quantum
\revision{Fisher-information} matrix.

Focused though we are on unitary processes, realize that arbitrary processes can be included by employing the same reasoning to develop a process-dependent norm optimized over general (including mixed) initial states $\rho$.
Not required for unitary processes, yet natural in developing the process norm is to generalize the optimization of states and measurements to be over an extended Hilbert space that includes ancillas in addition to the original system, even though the process itself acts only on the original system.
\revision{Sufficient it is to consider ancillary Hilbert spaces with dimension equal to that of the original system.}
The parametrized family of final states becomes $\rho_\coords=(\mathcal{I}\otimes\mathcal{E}_\coords)(\rho)$.
Thus define the process norm,
\begin{align}\label{eq:process-norm}
  \Vert\bvec{b}\Vert_{\mathcal{E}_\coords}^2
  =\max_\rho\,Q_{bb}
  =\max_\rho\,\tr\big(\rho L_b^2\big)
  =\max_\rho\,\max_{\text{measurements}|\rho}F_{bb}\,,
\end{align}
and generalize the chain~(\ref{eq:singlebound3}) of inequalities to \revision{a quantum-process bound,}
\begin{align}\label{eq:singlebound4}
\var{\hat\phi}\ge\frac{1}{F_{bb}}\ge\frac{1}{Q_{bb}}\ge\frac{1}{\Vert\bvec{b}\Vert_{\mathcal{E}_\coords}^2}\,.
\end{align}
More detail for this norm---indeed, that it is a norm---comes in App.~\ref{app:proc-norm}.

Easy it is to imagine that ancillas permit joint measurements that can extract
more information about the parameters, thus increasing the quantum Fisher information~$Q_{bb}$.
Indeed, the implicit definition~(\ref{eq:SLD}) of the SLD indicates that $L_b$ generally changes
when one allows joint system-ancilla states $\rho$.  Nonetheless, simple it is to argue that for
a unitary process $\mathcal{U}_\coords$, as in Eq.~(\ref{eq:unitaryprocess}), the process norm
is
\begin{align}\label{eq:unitaryprocessnorm}
\Vert\bvec{b}\Vert_{\mathcal{E}_\coords}=\Vert\bvec{b}H(\coords)\Vert_s\,,
\end{align}
even after including ancillas.
Suppose the maximum~(\ref{eq:process-norm}) for
a unitary process occurs on a mixed state $\rho$.
Purify $\rho$ into further ancillas, and find that the maximum occurs on a pure state.
Given that, run through the argument leading from  Eq.~(\ref{eq:Hphi}) to Eq.~(\ref{eq:seminorm}), and conclude with the result~(\ref{eq:unitaryprocessnorm}) for the unitary process norm.
\revision{Appreciate also that an argument from the convexity of the Fisher information demonstrates the optimality of pure states.\cite{fujiwara_quantum_2001}}

At first blush, nothing is gained for unitary processes by including ancillas
in the definition of the process norm.  On second look, however, there is
a there there.  Any extension to ancillas introduces degeneracies in the largest
and smallest eigenvalues of the generator~$\Id\otimes\bvec{b}H(\coords)$.
Degeneracies give more possibilities for the optimal fiducial state, an equal
superposition of states with largest and smallest eigenvalues; each such superposition
has its own associated optimal measurement.  These possibilities
can be put to use in probabilistic protocols, \revision{known to the optimal-design-of-experiments community as continuous designs,\cite{gazit_quantum_2019}} which flip a coin to choose among
different possibilities.
Classical the coin can be, or quantum by encoding
the coin into an entangled state.

In contrast to the universal expression for the process norm~(\ref{eq:unitaryprocessnorm})
for unitary processes, difficult it can be to evaluate the process norm for arbitrary processes.\cite{fujiwara_fibre_2008,demkowicz-dobrzanski_elusive_2012,kolodynski_efficient_2013,tsang_quantum_2013,alipour_quantum_2014,escher_general_2011}
Examples and explicit constructions in the remainder of this paper
specialize to unitary processes, but the derived bounds and their achievability
are applicable to arbitrary processes.

Now to one from many.  Estimation of $q$, as explained in
Sec.~\ref{sec:scalar-not-single-param}, is not the same as
the single-parameter problem of estimating $\phi=q$.  A fixed Fisher
information, we saw in Sec.~\ref{sec:classical-interpretation},
allows identification of a special single-parameter problem that gives the
appropriate one-from-many classical bound.  Discovering an analogous
single-parameter problem in the quantum setting, where the process norm
replaces the fixed Fisher information, is the subject of the
next subsection.

\begin{figure}
\centering
\includegraphics[width=0.50000\textwidth]{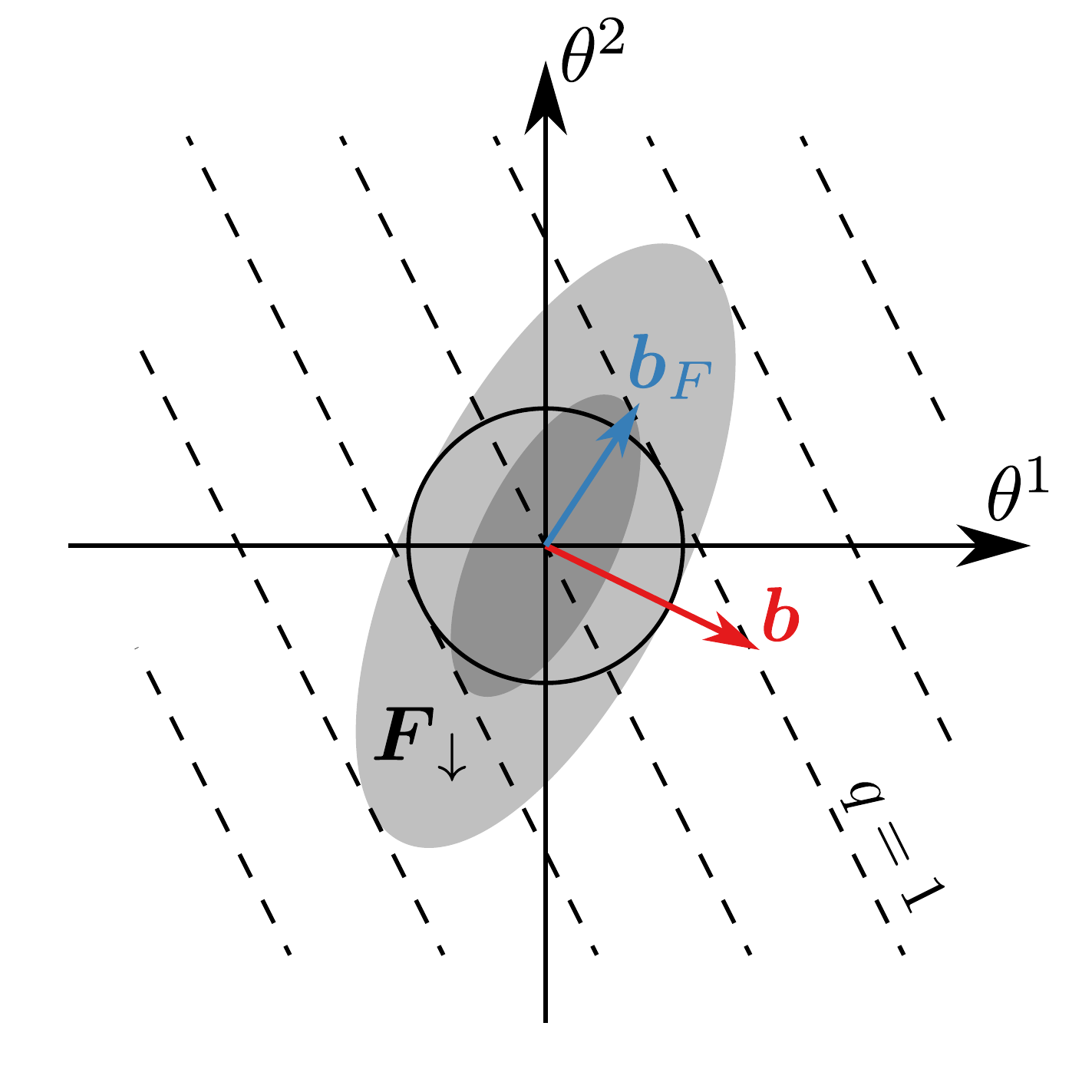}
\caption{For typical choices of $\bvec{b}$, saturating the \QCRB\ leaves the
one-from-many (Cauchy-Schwarz) inequality unsaturated:
$\bvec{F}_\downarrow(\bvec{b}_F,\bvec{b}_F)<F_{bb}=\Vert\bvec{b}\Vert_{\mathcal{E}_\coords}^2$.
The black circle represents the \QCRB, demarking the minimum width of a Fisher
ellipsoid in all directions.  The light-gray ellipse represents the CCRB on the
multiparameter estimator covariance given by a particular
preparation/measurement protocol.  Maximized by this protocol is the
single-parameter Fisher information $F_{bb}$, thus making
$F_{bb}=Q_{bb}=\Vert\bvec{b}\Vert_{\mathcal{E}_\coords}^2$, since the covariance ellipse
touches the quantum Cram{\'e}r-Rao circle along the direction $\bvec{b}$.  But
the shortest vector, according to $\bvec{F}_\downarrow$, that extends one unit
in $q$, is $\bvec{b}_F$, not $\bvec{b}$.  Failure to saturate the one-from-many
inequality is the result: $\bvec{F}(\bvec{b}_F,\bvec{b}_F)<F_{bb}$.
\label{fig:qcr}}
\end{figure}

\subsection{Scalar estimation}
\label{sec:quant-scalar-est}

Fuse the classical one-from-many bound~(\ref{eq:one-from-many-bound2})
with the single-parameter quantum bound~(\ref{eq:singlebound4}) to
obtain the ultimate chain, \revision{culminating in the quantum-process bound}:
\begin{align}\label{eq:ultimatechain}
\var{\hat q}
=\bvec{C}^\uparrow(dq,dq)
\ge\bvec{F}^\uparrow(dq,dq)
=\frac{1}{\bvec{F}_\downarrow(\bvec{b}_F,\bvec{b}_F)}
\ge\frac{1}{F_{bb}}
\ge\frac{1}{Q_{bb}}
\ge\frac{1}{\Vert\bvec{b}\Vert_{\mathcal{E}_\coords}^2}\,.
\end{align}
\revision{Pause yet again, this time to appreciate the interdependence of the quantities in the chain: the vector $\bvec{b}$ is partially constrained by the property $q$, the Fisher-information matrix $F$ is completely determined by the measurement protocol made up of an initial state and final measurement, and the vector $\bvec{b}_F$ is completely determined by combination of the property $q$ and the measurement protocol.}

Short-circuit from now on the first link in this chain, our interest being to work in
terms of classical Fisher ellipsoids, and \revision{also} the link through the quantum Fisher information
$Q_{bb}$, our quantum interest being to go directly to the ultimate bound of
optimized initial state, in which case
$Q_{bb}=\Vert\bvec{b}\Vert_{\mathcal{E}_\coords}^2$.  Upside-down turn the
chain~(\ref{eq:ultimatechain}), to work for convenience with Fisher informations:
\begin{align}\label{eq:QCRB2}
\bvec{F}_\downarrow(\bvec{b}_F,\bvec{b}_F)\le F_{bb}\le
\Vert\bvec{b}\Vert_{\mathcal{E}_\coords}^2\,.
\end{align}
The first inequality is one-from-many (or Cauchy-Schwarz); the second, the
\QCRB.  Pose now a new question: for what $\bvec{b}$ are these two
inequalities saturated?  More precisely, is there a single-parameter problem
defined by $\bvec{b}$, $dq(\bvec{b})=1$, with an optimal estimation protocol
(saturating the \QCRB\ inequality) where errors in estimates of the other parameters that
are irrelevant to $q$ are uncorrelated with errors in the single parameter
associated with $\bvec{b}$ (saturating the one-from-many inequality)?

Available already is the condition for saturating the one-from-many inequality:
The tip of $\bvec{b}$ must be at a point of tangency between the unit surface of
$q$ and a Fisher covariance ellipsoid of some measurement $\bvec{F}_\downarrow$.

Developing a similar geometric picture for the second inequality is the task
now.  To the fore comes the norm defined by the \QCRB,
$\Vert\bvec{b}\Vert_{\mathcal{E}_\coords}^2$, and as the geometric object of
interest, the ``circle'' of vectors of constant \QCRB\ norm.  Before
optimization over initial states, the quantum Fisher-information matrix is a
quadratic form, whose ``circles'' of constant norm are ellipsoids.  As $\bvec b$
changes direction, however, the optimal initial state changes.  The reason for
this change is clear in the unitary case, where changing the generator
$\bvec{b}H(\coords)$ changes the eigenstates with extremal eigenvalues from
which the optimal state is built.  The result is to give the \QCRB\ norm more
diverse, nonmetric unit-circle shapes, even shapes with corners.
Draw surfaces of constant \QCRB\ norm, which are free of the dependence
on initial state that plagues surfaces of constant QCRB, always remembering
that to saturate the \QCRB\ as a single-parameter estimation problem for a
particular $\bvec{b}$ requires using an optimal initial state for $\bvec{b}$
and making the corresponding optimal measurement.

\revision{Crucial} it is to note that the \QCRB\ unit ``ball,'' \ie, the circle
of unit process norm and the interior of the circle, as the intersection of the Fisher
ellipsoids for all measurements and all fiducial states, is an (absolutely) convex set.
More technical discussion of the process norm and its unit ball is given in
App.~\ref{app:proc-norm}.

\begin{figure}
\centering
\includegraphics[width=\textwidth]{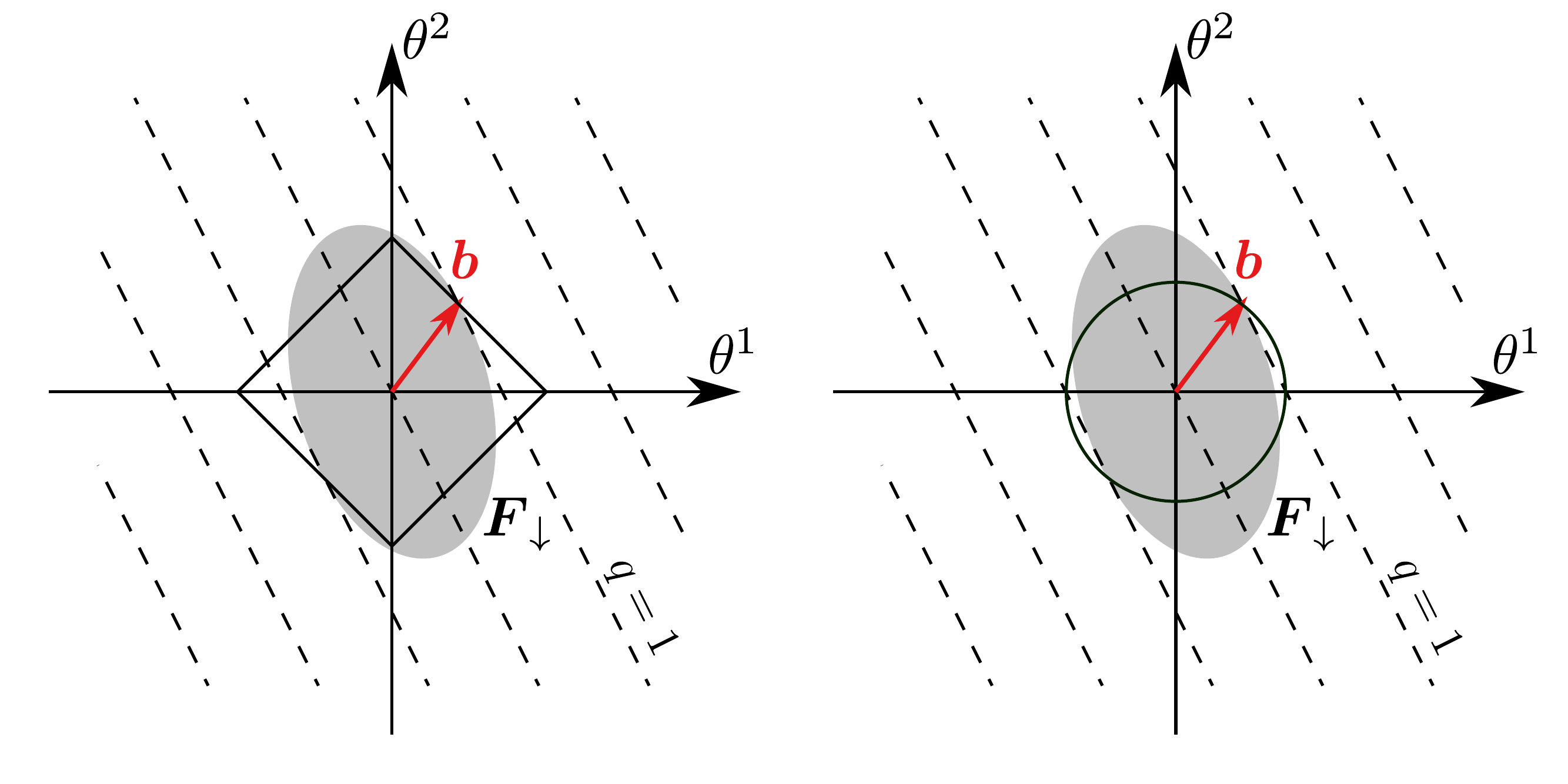}
\caption{If the surface of constant \QCRB\ passing through the tip of $\bvec{b}$
also passes through the unit surface of $q$, impossible it is for
$\bvec{b}=\bvec{b}_F$, since this would imply the Fisher ellipsoid enters the
interior of the surface of constant \QCRB\ norm.  Illustrated here is how
this works for a nonmetric (square) \QCRB\ surface (left) and a metric (circular)
\QCRB\ surface (right).}
\label{fig:non-kissing-qcrbs}
\end{figure}

\begin{figure}
\centering
\includegraphics[width=\textwidth]{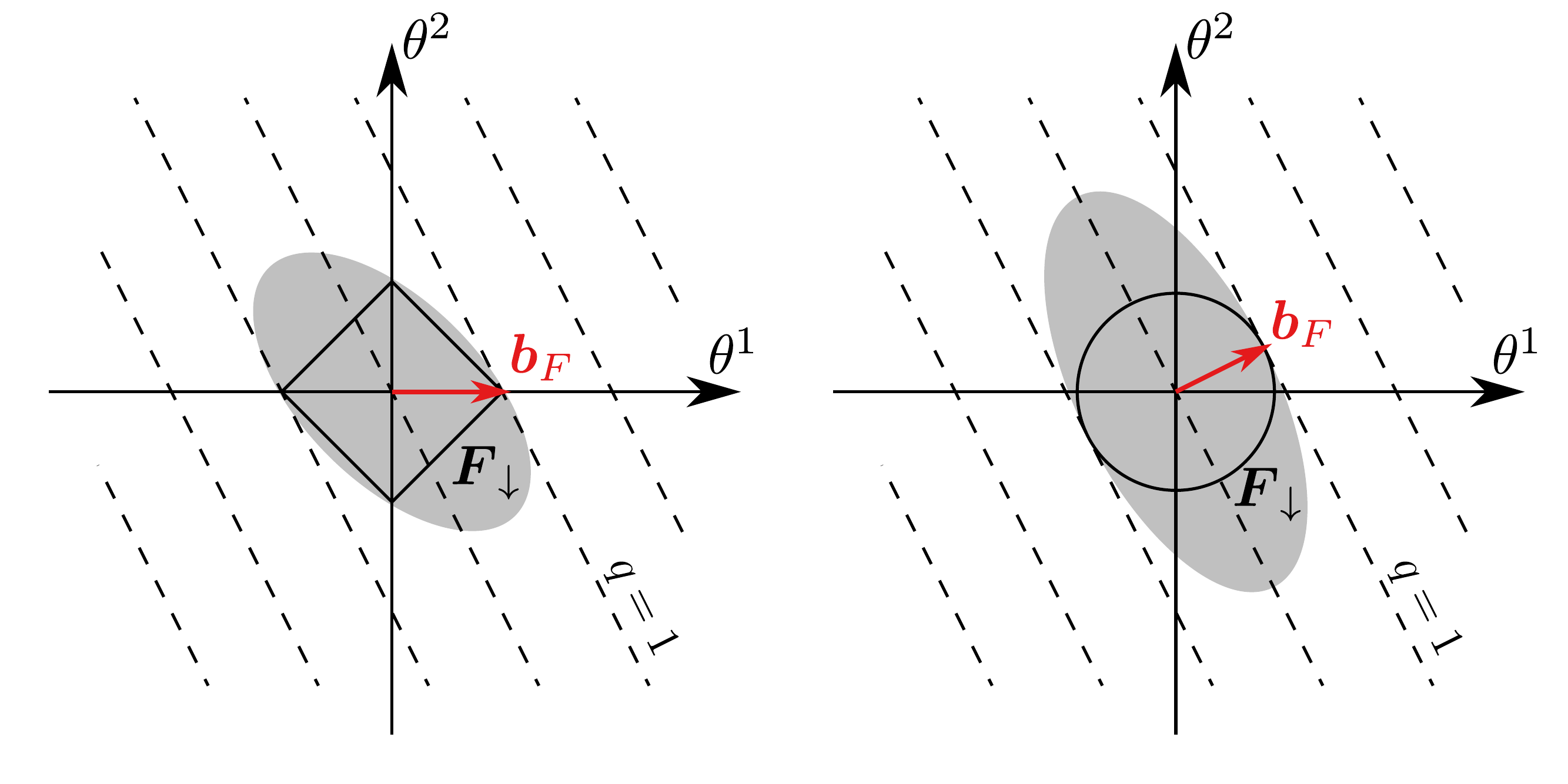}
\caption{If the surface of constant \QCRB\ passing through the tip of $\bvec{b}$
does not pass through the unit surface of $q$, one can draw an ellipsoid tangent
to the level surfaces of $q$ at the tip of $\bvec{b}$ that remains outside the \QCRB\ surface.
This ellipsoid corresponds to a Fisher information that makes $\bvec{b}=\bvec{b}_F$.  Illustrated here is how
this works for a nonmetric (square) \QCRB\ surface (left) and a metric (circular)
\QCRB\ surface (right).  The vector $\bvec{b}$ is the shortest vector satisfying $dq(\bvec{b})=1$ according
to the \QCRB\ norm.}
\label{fig:kissing-qcrbs}
\end{figure}

A single-parameter problem must simultaneously saturate the one-from-many
inequality and the \QCRB\ in order to qualify as no-control estimation.
As illustrated in Fig.~\ref{fig:qcr}, saturating the \QCRB\ does not guarantee
that the one-from-many inequality is saturated, because an initial
state and measurement procedure that saturate the \QCRB\ for some $\bvec{b}$
generally has a vector $\bvec{b}_F$, the shortest vector according to
the classical Fisher metric for this state and measurement, that is
different from $\bvec{b}$.  Likewise, saturating the one-from-many inequality
does not guarantee saturation of the \QCRB.

To find conditions for the two vectors to coincide, draw a vector $\bvec{b}$
satisfying $dq(\bvec{b})=1$ and then the surface of constant \QCRB\ norm that
passes through the tip of $\bvec{b}$.  The optimal initial state and optimal
measurement that attain the \QCRB\ for $\bvec{b}$ have a Fisher ellipsoid,
$F_{vv}=F_{bb}=\Vert\bvec{b}\Vert_{\mathcal{E}_\coords}^2$, that passes through
the tip of $\bvec{b}$; the QCRB~(\ref{eq:QCRB2}) implies that this Fisher
ellipsoid cannot enter the interior of the surface of constant \QCRB\ norm.  But
if we are to have $\bvec{b}=\bvec{b}_F$ for this  measurement, then the Fisher
ellipsoid is tangent to the unit surface of $q$ at $\bvec b=\bvec b_F$.

Close in on the quarry: conclude that the surface of constant \QCRB\ norm cannot
pass through the unit surface of $q$ as in Fig.~\ref{fig:non-kissing-qcrbs}; it
must kiss that surface as in Fig.~\ref{fig:kissing-qcrbs}.  If the surfaces of
constant \QCRB\ norm are smooth, this kissing is tangency; if not, only kissing.
Now the kill: conclude further, using the argument used for Fisher ellipsoids,
that $\bvec{b}$ is the shortest vector $\bmin$, according to the \QCRB\ norm, whose
tip lies on the unit surface of~$q$,
\begin{align}
  \bvec{b}_{\mathrm{min}}
  =\arg\min_{\bvec{b}}\big\{\Vert\bvec{b}\Vert_{\mathcal{E}_\coords}^2\,\big\vert\,dq(\bvec{b})=1\big\}\,.
\end{align}

\begin{figure}
\centering
\includegraphics[width=1.00000\textwidth]{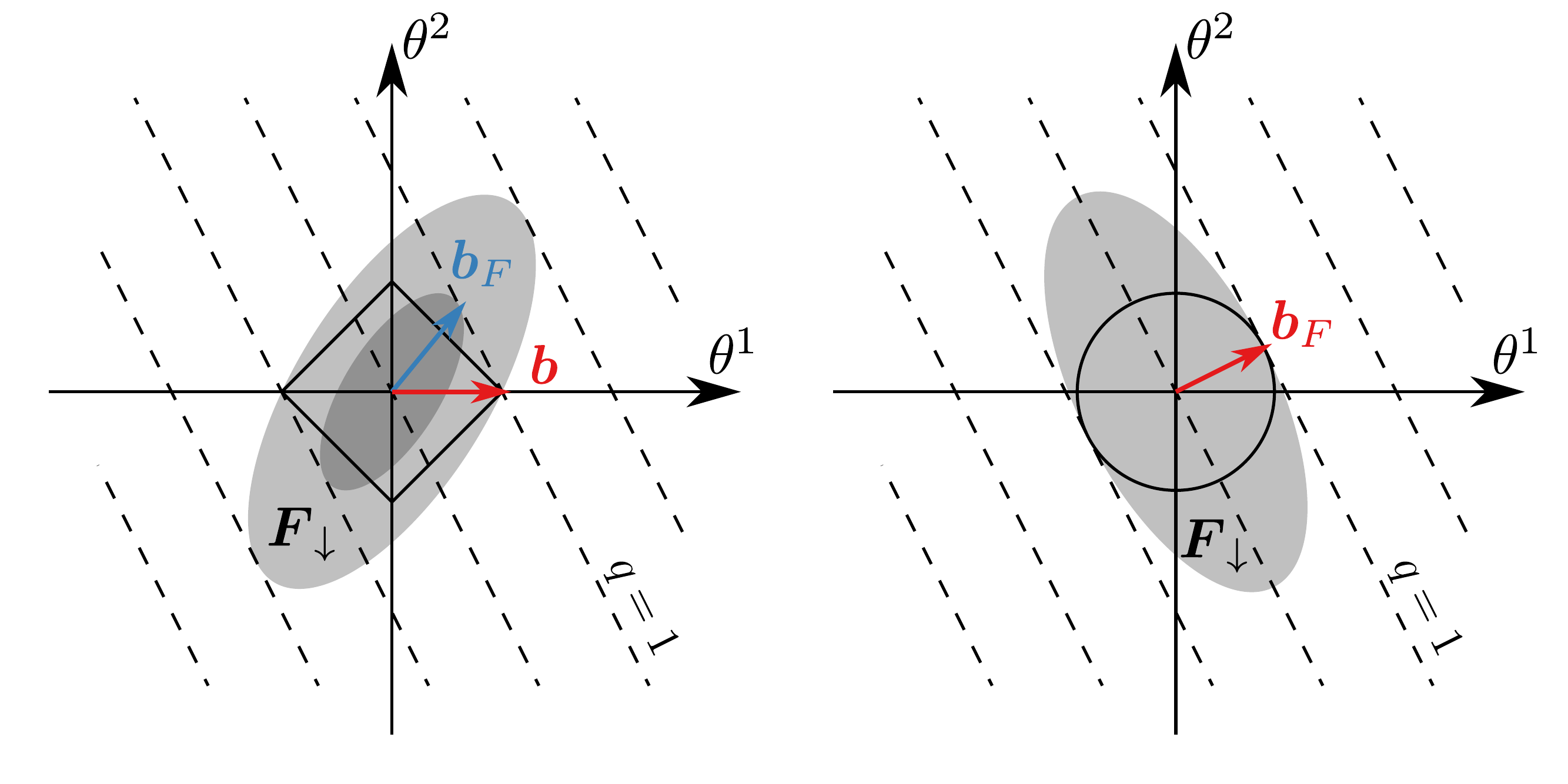}
\caption{For smooth \QCRB\ unit circles (illustrated at right), saturating
covariance ellipses have a unique tangent to the unit circle that must
match the level surfaces of $dq$. At corners (illustrated at left),
there are many tangents that a saturating covariance ellipse might make,
so it is not guaranteed to be tangent to the level surfaces of
\(dq\).\label{fig:max}}
\end{figure}

\begin{figure}
\centering
\includegraphics[width=0.50000\textwidth]{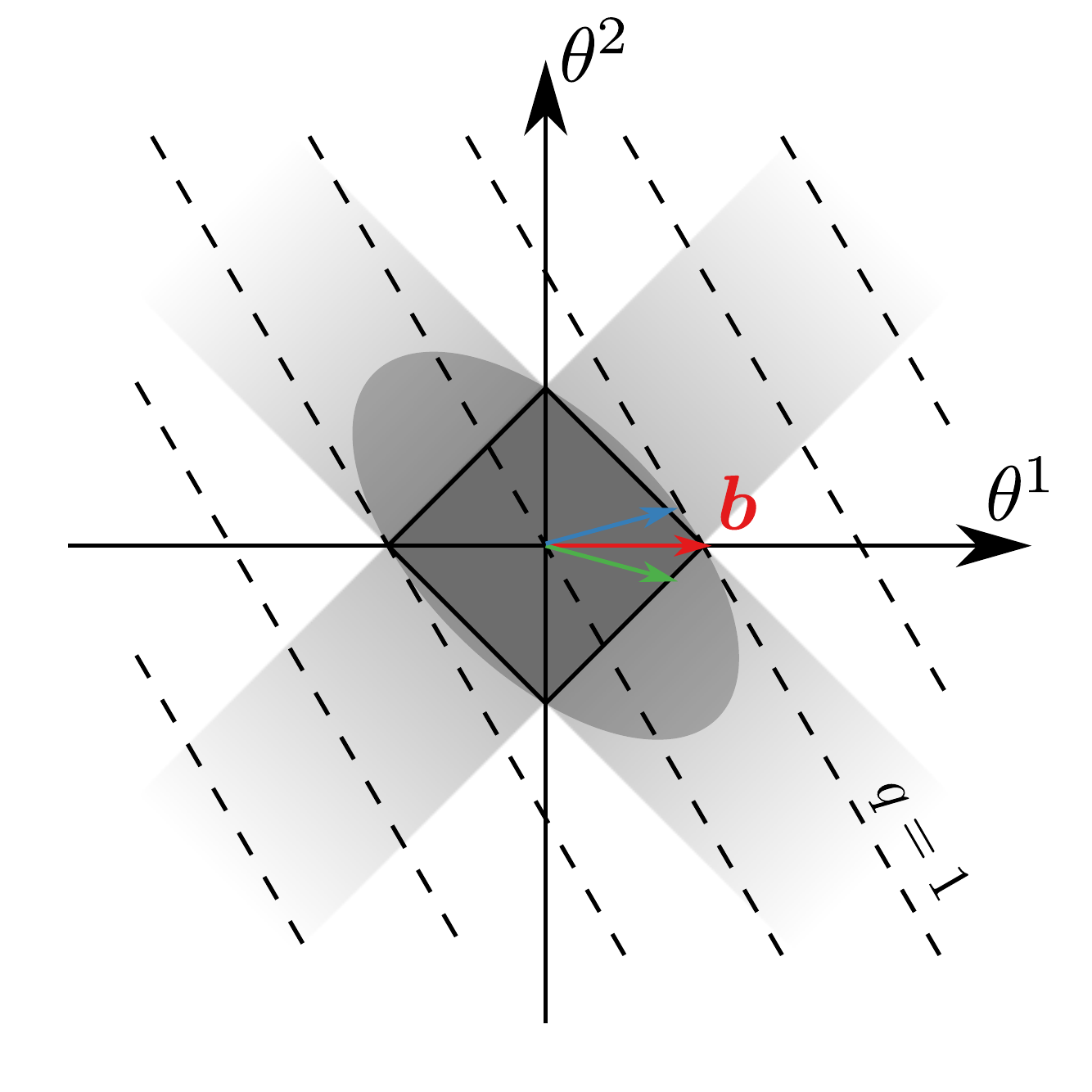}
\caption{When the shortest \(\bvec{b}\) according to the \QCRB\
norm lies on a corner of the unit circle, measurements that saturate the quantum
\QCRB\ bound for nearby parameters (the blue and green arrows) also
saturate the \QCRB\ bound for \(\bvec{b}\). These covariance
ellipses have infinite extent in parameters tangent to the unit circle at these
nearby points, so the tangents they make to the corner are the most extreme
possible.  Flipping a coin to choose between protocols that saturate the
quantum Cram{\'e}r--Rao bounds for these nearby parameters leads to covariance
ellipses that still saturate the quantum Cram{\'e}r--Rao bound for \(\bvec{b}\)
while making a tangent to the corner that interpolates between the extreme
tangents.  Choose therefore the probabilities to create a tangent that matches
the level surfaces of $dq$, simultaneously saturating the one-from-many
inequality and the \QCRB~bound.\label{fig:sat-corner}}
\end{figure}

One question yet remains: is it always possible to saturate the one-from-many
(Cauchy-Schwarz) inequality by discovering a state preparation and measurement
protocol such that  $\bvec{b}_F=\bmin$?  One cannot
fail if the Cram{\'e}r--Rao unit circle is smooth at $\bmin$.  As illustrated
in Fig.~\ref{fig:max}, all measurements saturating \QCRB\ for a smooth unit
circle have a Fisher \hbox{ellipsoid} that is tangent to the unit level surface of $q$,
thus also saturating one-from-many.  Care is required if the unit circle is
pointed at $\bmin$: \QCRB-saturating protocols generally have measurements with
Fisher ellipsoids that pass through the $q=1$ surface, thus failing to saturate
one-from-many, but measurements that do the job can be constructed.  In
particular, probabilistic measurement procedures transform extremal
\QCRB-saturating protocols into protocols that also saturate the one-from-many
inequality, as illustrated in Fig.~\ref{fig:sat-corner} and demonstrated
in App.~\ref{app:convex-fisher}.

Contemplate the happy situation: the Fisher ellipsoid for the measurement-state
combination and the surface of constant \QCRB\ norm both kiss the unit surface
of $q$ at $\bmin$, with the Fisher ellipsoid lying (inclusively) between the
\QCRB\ surface and the unit surface of~$q$.  Understood is that tangency and
kissing include the case where surfaces coincide; then ``shortest''  means ``no
shorter,'' and a kiss generalizes to a more generous smooch across a portion of
a planar surface.

A Fisher ellipsoid has a tangent plane that kisses the unit surfaces of $q$ at $\bmin$ if and only if
(i)~$\bmin=\bvec{b}_F$, \ie, all $\bvec{v}_\perp$ that don't change $q$, are orthogonal to $\bmin$ according
to the Fisher metric,
\begin{align}\label{eq:kiss1}
  \bvec{F}_\downarrow(\bmin,\bvec{v}_\perp)
  &=0=\Vert\bmin\Vert_{\mathcal{E}_\coords}^2dq(\bvec{v}_\perp)\,,
\end{align}
and (ii)~the Fisher information therefore saturates both inequalities in~(\ref{eq:QCRB2}),
\begin{align}\label{eq:kiss2}
  F_{bb}
  &=
  \bvec{F}_\downarrow(\bmin,\bmin)
  =
  \Vert\bmin\Vert_{\mathcal{E}_\coords}^2
  =
  \Vert\bmin\Vert_{\mathcal{E}_\coords}^2dq(\bmin)
  \,.
\end{align}
Combining Eqs.~(\ref{eq:kiss1}) and~(\ref{eq:kiss2}) gives a single unified kissing condition \revision{for achieving the quantum-process bound~(\ref{eq:ultimatechain})},
\begin{align}\label{eq:kissing-condition1}
  \bvec{F}_\downarrow(\bmin,\bvec{v})
  =
  \Vert\bmin\Vert_{\mathcal{E}_\coords}^2dq(\bvec{v})\quad\mbox{for all $\bvec{v}$}\,,
\end{align}
which is equivalent to
\begin{align}\label{eq:kissing-condition2}
  F_{jk}b_{\mathrm{min}}^j=\Vert\bmin\Vert_{\mathcal{E}_\coords}^2q_k
  \quad\Longleftrightarrow\quad
  b_{\mathrm{min}}^j=\Vert\bmin\Vert_{\mathcal{E}_\coords}^2q^j\,.
\end{align}

\revision{Questions of saturability having been laid to rest, present now a particularly appealing way of writing the final, attainable quantum-process bound on estimating a scalar function of a multiparameter quantum process,
\begin{align}
\var{\hat q}
\ge\frac{1}{\Vert\bmin\Vert_{\mathcal{E}_\coords}^2}
=\Vert dq\Vert_{\mathcal{E}_\coords *}^2\,.
\label{eq:main-result}
\end{align}
The last equality employs the norm $\Vert\cdot\Vert_{\mathcal{E}_\coords *}$ dual to the process norm to express the bound entirely in terms of the process family and property to be estimated.  Results analogous to Eq.~(\ref{eq:main-result}) in the case of parameterized families of quantum states are known.\cite{suzuki_nuisance_2019,tsang_quantum_2020}  In such a case, where the process norm arises from a quadratic form $\bvec{Q}_\downarrow$, the dual norm is naturally expressed in terms of the inverse $\bvec{Q}^\uparrow$: $\var{\hat q}\ge\bvec{Q}^\uparrow(dq,dq)=1/\bvec{Q}_\downarrow(\bmin,\bmin)$.}

\revision{Appropriate it is to comment on dependence of the optimal estimation strategy on the fiducial state.  Appreciate that in general there does not exist a global estimation strategy that is optimal for all fiducial states.  Inconvenient this is, indeed inconvenient enough to prompt consideration of the local problem at hand.}

Explore this answer through examples in Sec.~\ref{sec:examples}, but before doing so,
clarify in the next subsection that there are other ways to perform what might be called
no-control estimation and how these are related to the results in this paper.

\subsection{Intervention techniques}

Easy it is to imagine protocols
that rescale the parameters in the unitary operator
$e^{-iH(\coords)}=e^{-i\theta^j X_j}$ by changing the constants that couple
a generator to the system or, equivalently, by adjusting separately the
evolution times for those generators.  Spin echo can accomplish this
effect without directly adjusting coupling constants or evolution times.
Such rescaling effectively changes the Hamiltonian, yet might be
regarded as a no-control protocol, since rather than directly controlling an
underlying parameter in the Hamiltonian, the protocol controls quantities,
associated with a generator, that are generally available to an agent
in charge of a metrological experiment.  Appreciating this argument, nonetheless
we stick with the approach outlined up till now: the family of
processes $\mathcal{E}_\coords$ is part of the statement of the
problem---completely specified by the Hamiltonian $H(\coords)=\theta^jX_j$
for unitary processes; separate scaling of the parameters via intervention
techniques leads to a problem that, though readily analyzed by the techniques
developed in this paper, is nonetheless a different problem.  Tying the
notion of a parameter to the process family and sticking with that notion
fixes the method by which the parameters are impressed on the system,
enabling us to extract a magic number from the process family, the ultimate
quantum limit given by the square of the process norm
$\Vert\bmin\Vert_{\mathcal{E}_\coords}^2$,
which for unitary processes becomes the squared seminorm of the generator,
$\Vert\bmin H(\coords)\Vert_s^2$.

\section{Examples: Putting the formalism to work}
\label{sec:examples}

\subsection{Commuting generators}
\label{sec:commuting-generators}

\subsubsection{Setup}
\label{sec:commuting-setup}

Consider now the scenario introduced by Eldredge~{\it et
al.}\cite{ref-eldredge_optimal_2016}: the parameters $\theta^j$ are rotation
angles about the Bloch $z$ axis for different qubits; hence, the generators are
Pauli $z$ operators $\sigma_j^z$ for the various qubits, giving Hamiltonian
\begin{align}
H(\tilde\theta)&=\tfrac12\theta^j\sigma^z_j\,.
\end{align}
For convenience and without any loss of generality, discard qubits that
do not contribute to $q$, \ie, for which $q_j=0$; order the remaining qubits
so that the absolute value of $q_j$ descends through the list of qubits;
and scale $q$ such that $q_1=1$, thus giving
$1=q_1\geq|q_2|\geq\cdots\geq|q_N|>0$.

For an arbitrary vector $\bvec{b}=b^j\partial_j$, the single-parameter generator and \QCRB\ norm are
\begin{align}\label{eq:Ycommuting}
\Gen&=\bvec{b}H(\tilde\theta)=\frac12 b^j\sigma^z_j\,,\\
\label{eq:QCRBcommuting}
\Vert\bvec{b}H(\tilde\theta)\Vert_s&=
\frac{1}{2}\Vert b^j\sigma^z_j\Vert_s=\sum_{j=1}^N|b^j|=\Vert\bvec b\Vert_1\,.
\end{align}
Here $\Vert\bvec b\Vert_1$ is the 1-norm of the vector $\bvec b$.

The geometric object of interest is the \QCRB\ unit surface, $\Vert\bvec
b\Vert_1=1$.  This, the unit cross-polytope in $N$ dimensions, is the dual of
the unit hypercube.  In three dimensions, the cross-polytope is the octahedron.
A hyperface of the cross-polytope lies in the unit plane defined by a linear
function $z=z_j\theta^j$, with $z_j=\pm1$.  Indeed, a hyperface is the
intersection of the cross-polytope with the unit plane of $z$,
\begin{align}
dz(\bvec b)=\bvec b z=z_jb^j=1\,.
\end{align}
Here $dz=z_jd\theta^j$ is the 1-form corresponding to $z$.  Thus define a hyperface of the
cross-polytope by
\begin{align}\label{eq:hyperface}
\left\{\bvec b\,\left|\,z_jb^j=1
\quad\mbox{and}\quad
\sum_{j=1}^N|b^j|=1\right.\right\}\,.
\end{align}
Stress that the sign of $b^j$ is $z_j$, implying that $b^j=z_j|b^j|$ (no sum).
Convenient and productive it is to let a string $\tz=z_1\ldots z_N$ list the
coefficients $z_j$ and so specify a hyperface.

Construct first in Sec.~\ref{sec:hyperface} measurements optimal for a $q$ that
coincides with a hyperface $\tz$, \ie, $dq=dz$; then use these measurements in
Sec.~\ref{sec:kiss} as ingredients in the recipe for measurements optimal at the
\QCRB\ corners that arise for general $q$.

\subsubsection{Hyperface measurements}
\label{sec:hyperface}

When the unit surface of $q$ coincides with a hyperface $\tz$ there are many
choices for $\bmin$ (any vector in the hyperface $\tz$ will do).  Aesthetic
sensibilities direct us to $\bmin$ with components $b^j=1/z_jN$.  The
generator associated with this choice,
\begin{align}
  Y=\bmin H(\coords)=\frac{1}{2}\frac{1}{z_jN}\sigma_j^z\,,
\end{align}
has extremal eigenvalues $\pm\frac{1}{2}$ associated with the eigenvectors
\begin{align}
  |\pmtz\rangle=\bigotimes_{j=1}^N|{\pm}z_j\rangle\,.
\end{align}
Here $|z_j\rangle$ is the eigenstate of $\sigma^z_j$ with eigenvalue $z_j$,
\ie, $\sigma_j^z|z_j\rangle=z_j|z_j\rangle$, and $\minustz$ is the string with
the sign of all the entries reversed, \ie, ${\minustz}=-z_1\,\ldots{-z_N}$;
$\tz$ and $\minustz$ specify opposite faces of the cross-polytope.  The normalized
states $|\tz\rangle$ are orthogonal:
\begin{align}
\langle\tz|\tz'\rangle=\prod_{j=1}^N\langle z_j|z'_j\rangle=\delta_{\tz\tz'}\,.
\end{align}

Define cat-superposition states,
\begin{align}\label{eq:cat}
|\psi_\tz^{(\pm)}\rangle&=\frac{1}{\sqrt2}\big(|\tz\rangle\pm|{\minustz}\rangle\big)\qquad\mbox{(cat state),}\\
|\psi_\tz^{(\pm i)}\rangle&=\frac{1}{\sqrt2}\big(|\tz\rangle\pm i|{\minustz}\rangle\big)\qquad\mbox{($i$cat state)}\,.
\label{eq:icat}
\end{align}
Any of these choices work as the initial state for an optimal estimation
strategy.  Choosing $|\psi_\tz^{(+)}\rangle$ and imposing the parameters via
$H(\coords)$ yields the final state
\begin{align}
e^{-iH(\tilde\theta)}|\psi_\tz^{(+)}\rangle
=\frac{1}{\sqrt2}\big(e^{-iz_j\theta^j/2}|\tz\rangle
+e^{iz_j\theta^j/2}|{\minustz}\rangle\big)\,.
\end{align}
Measure now in an orthonormal basis containing
$|\psi_{\tz}^{(\pm i)}\rangle$ (the basis elements in the subspace orthogonal
to the span of $|\tz\rangle$ and $|\minustz\rangle$  are superfluous, since the
final state has no support on that subspace).  Appendix~\ref{app:comm-gen-details}
shows how to think of the needed measurement as a parity measurement and thus how to
implement it locally.

The probabilities for the results corresponding to states
$|\psi_{\tz}^{(\pm i)}\rangle$ are
\begin{align}
  p(\pm|\tilde\theta)
  =
  \big|\langle\psi_{\tz}^{(\pm i)}|e^{-iH(\coords)}
  |\psi_{\tz}^{(+)}\rangle\big|^2
  =
  \frac14\big|e^{-iz_j\theta^j/2}\mp i e^{iz_j\theta^j/2}\big|^2
  =
  \frac12\big(1\pm\sin z_j\theta^j\big)\,,
\end{align}
leading to Fisher-information matrix
\begin{align}\label{eq:Fjkz}
  F_{jk}
  =
  \sum_{\pm}\frac{1}{p(\pm|\coords=0)}
  \left.\frac{\partial p(\pm|\coords)}{\partial\theta^j}\right|_{\coords=0}
  \left.\frac{\partial p(\pm|\coords)}{\partial\theta^k}\right|_{\coords=0}
  =
  z_jz_k\,;
\end{align}
\ie, $\bvec{F}_\downarrow=dz\otimes dz$ is the degenerate \revision{Fisher-information} matrix for
no-control estimation of $q=z$ discussed at the end of Sec.~\ref{sec:classical-interpretation}.

\begin{figure}
\centering
\includegraphics[width=\textwidth]{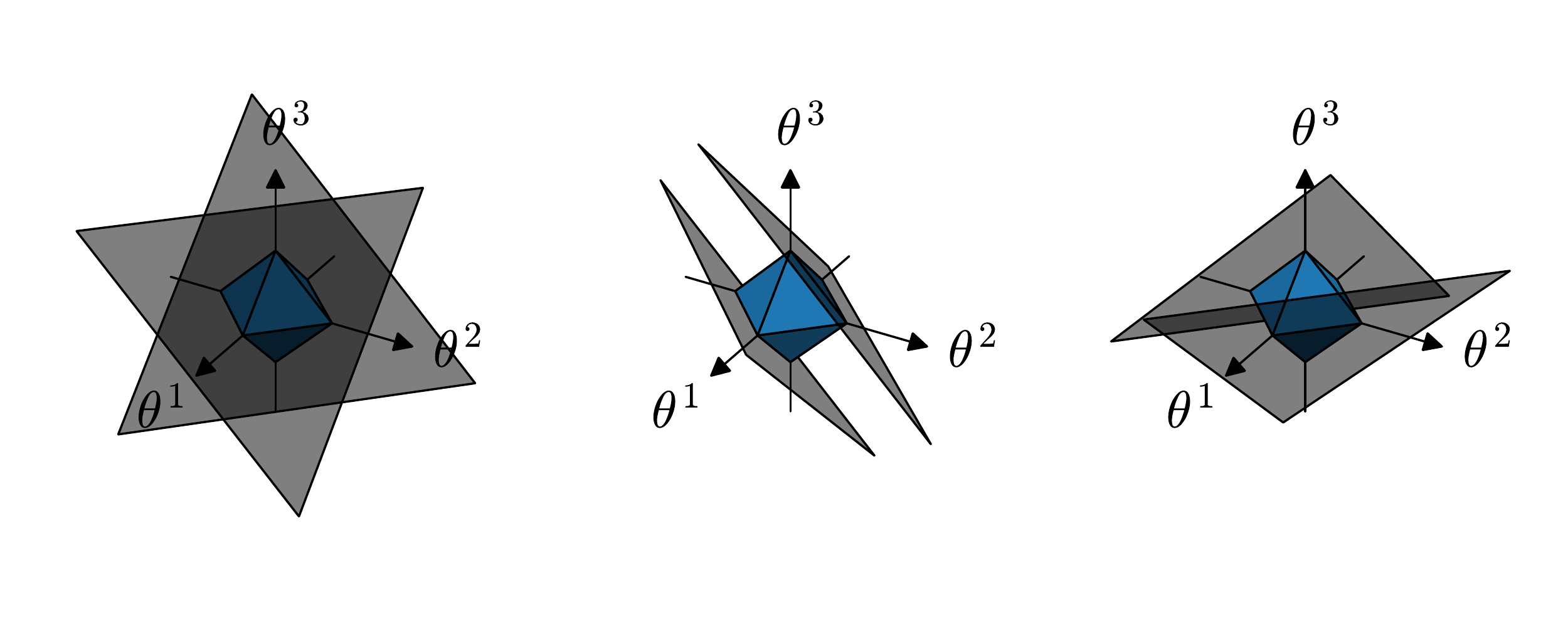}
\caption{A sampling of Fisher ``ellipsoids'' for hyperface measurements in the
case $N=3$, where the \QCRB\ unit surface, the $N=3$ cross-polytope, is the (blue)
octahedron.  The hyperfaces are depicted by pairs of shaded triangles
bounding the covariance regions.  The chosen samples correspond to the
Fisher informations $\bvec{F}_\downarrow^{(k)}$ used to saturate \QCRB\ when
all $q_j>0$.  The corresponding $\tz$ strings are, from left to right,
$\tone\tone\tone$, $\tone\tminusone\tminusone$, and $\tone\tone\tminusone$.
\label{fig:extremal-faces}}
\end{figure}

Verify that the tangency condition~(\ref{eq:kissing-condition2}) is met for $q_j=z_j$, noting
that $\Vert\bmin H(\coords)\Vert_s=\Vert\bmin\Vert_1=1$:
\begin{align}
  F_{jk}b_{\mathrm{min}}^j=\sum_{j=1}^N z_jz_k\frac{1}{z_jN}
  =z_k=\Vert\bmin H(\coords)\Vert_s^2q_k\,.
\end{align}
Figure~\ref{fig:extremal-faces} illustrates the Fisher ``ellipsoids'' for
several of these optimal measurements in the case $N=3$.

\subsubsection{Getting away with kissing at corners}
\label{sec:kiss}

Arbitrary $q$, put in the canonical form described in
Sec.~\ref{sec:commuting-setup}, now comes to the fore.
Broken is the symmetry of $q=z$; the unit cross-polytope
is only guaranteed to touch the unit surface of $q$ at one point,
\begin{align}
  \bmin=\partial_1\,.
\end{align}
This vector lives at the corner of the unit circle just like the vector in
Fig.~\ref{fig:sat-corner}.  Implement the probabilistic corner strategy discussed
at the end of Sec.~\ref{sec:quant-scalar-est}: construct a Fisher
ellipsoid whose tangent surface matches the level surfaces of $q$ at
$\bmin$ (and hence saturates the one-from-many inequality) by using a
convex combination of Fisher informations saturating the \QCRB\ on the
hyperfaces adjacent to that corner.

\begin{figure}
  \centering
  \includegraphics[width=0.5\textwidth]{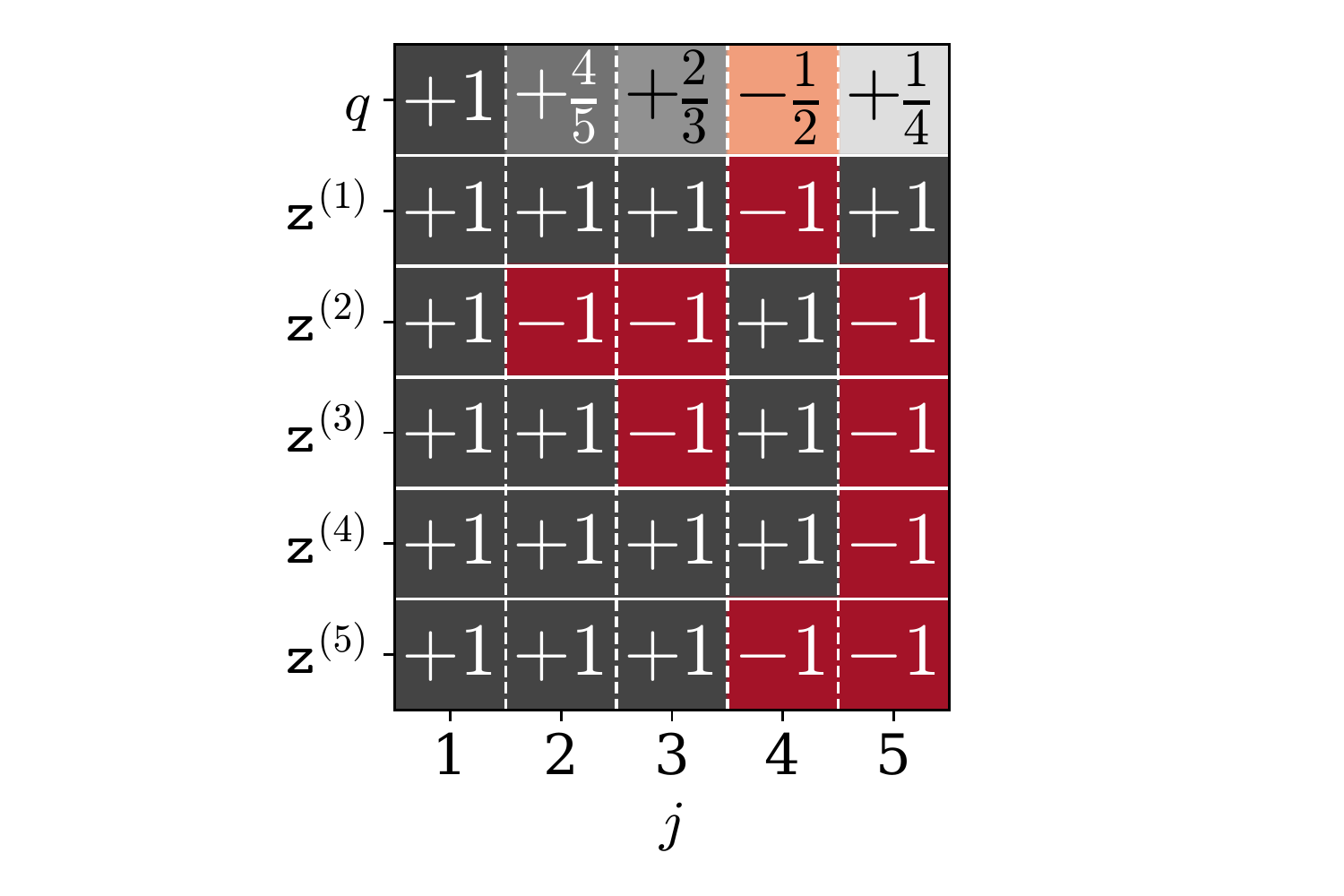}
  \caption{Illustration of the special strings $\tz^{(k)}$ for
$q=\theta^1+\frac{4}{5}\theta^2+\frac{2}{3}\theta^3-\frac{1}{2}\theta^2+\frac{1}{4}\theta^5$.
The parameters $\theta^j$ have been ordered such that $|q_j|\geq|q_k|$ for $k\geq j$.}
  \label{fig:special-z-strings}
\end{figure}

Define the strings $\tz^{(k)}$ corresponding to the adjacent hyperfaces,
\begin{align}
  z_j^{(1)}&=\sgn q_j=\begin{cases}1 & q_j>0 \\ -1 & q_j<0\end{cases}\,,\\
  z_j^{(k>1)}&=\begin{cases}z_j^{(1)} & \mbox{if }j<k\,,\\
  -z_j^{(1)} & \mbox{if }j\geq k\end{cases}\,,\\
  dz^{(k)}&=z_1^{(1)}d\theta^1+\cdots+z_{k-1}^{(1)}d\theta^{k-1}
  -z_k^{(1)}d\theta^k-\cdots-z_N^{(1)}d\theta^N\vphantom{\Bigg(}\,.
\end{align}
Figure~\ref{fig:special-z-strings} depicts these strings for a particular $q$
when $N=5$.  Appreciate now two important properties of these strings:
first, $\{dz^{(k)}\}$ is a basis of forms; second, the coefficients of $dq$
in this basis are positive and normalized to unity---they make up a probability
distribution.  Specifically,
\begin{align}
dq=\sum_{k=1}^N p_k dz^{(k)}\,,
\end{align}
with
\begin{align}
  p_1&=\frac{1}{2}\big(1+|q_N|\big)\,, \\
  p_{k>1}&=\frac{1}{2}\big(|q_{k-1}|-|q_k|\big)\,.
\end{align}

\begin{figure}
\centering
\includegraphics[width=0.5\textwidth]{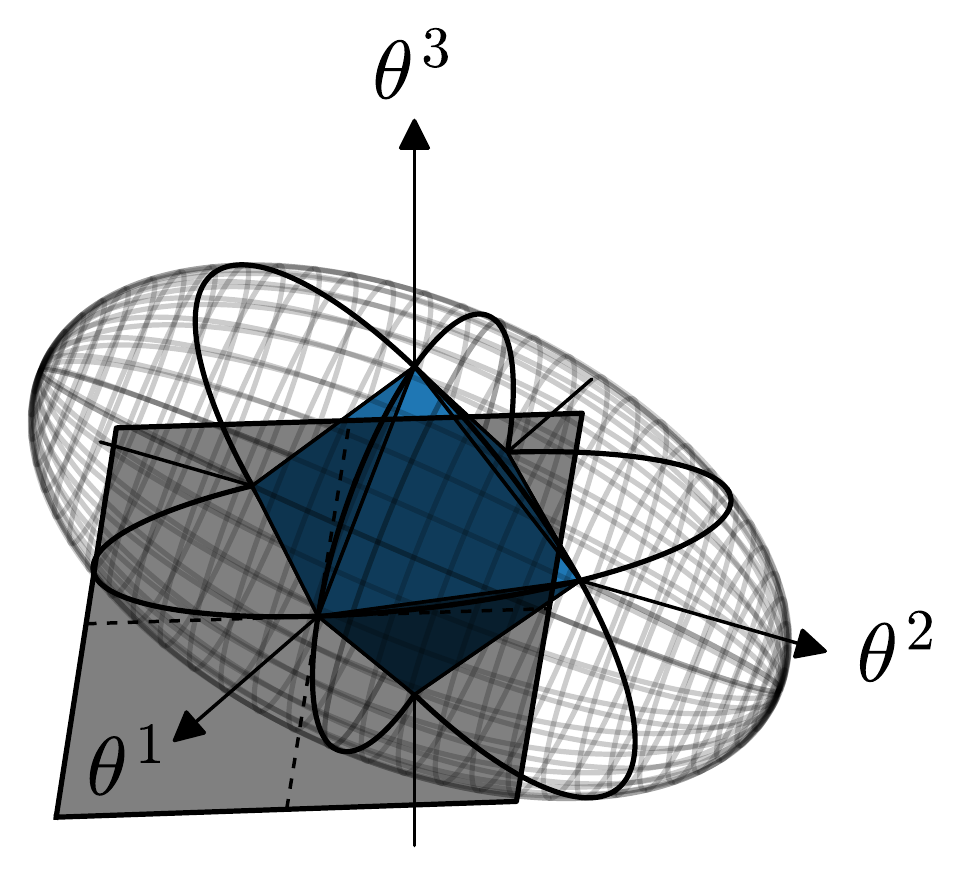}
\caption{The Fisher information (illustrated as a mesh ellipsoid) whose tangent
plane at $\bvec{b}_{\mathrm{min}}=\partial_1$ (illustrated as the shaded plane)
is a level surface of
$q=\theta^1+\frac{2}{3}\theta^2+\frac{1}{3}\theta^3$. This ellipsoid contacts
the \QCRB\ unit circle, $\Vert\bvec{b}\Vert_{\mathcal{U}_{\tilde{\theta}}}=\Vert\bvec{b}H(\tilde\theta)\Vert_s=1$,
illustrated as the blue octahedron, at all vertices.\label{fig:tangent-match}}
\end{figure}

The Fisher informations for the hyperface measurements are
$\bvec{F}^{(k)}_\downarrow=dz^{(k)}\otimes dz^{(k)}$.  Performing the
$\bvec{F}^{(k)}$ measurement with probability $p_k$ yields the new
Fisher information, $\bvec{F}_\downarrow=\sum_kp_k\bvec{F}_\downarrow^{(k)}$.
Appendix~\ref{app:convex-fisher} explains why such a protocol is allowed
and why the Fisher information takes this form.  Verify now that the kissing
condition~(\ref{eq:kissing-condition1}) is satisfied:
\begin{align}
\bvec{F}_\downarrow(\bmin,\bvec{v})
=\sum_{k=1}^N p_k dz^{(k)}(\bvec{v})
=\Vert\bvec{b}H(\tilde\theta)\Vert_s^2\,dq(\bvec{v})\,.
\end{align}
Fig.~\ref{fig:tangent-match} illustrates a Fisher information constructed
according to this recipe.

Similar in spirit is this construction to that in Sec.~IV\,B\,1 of Eldredge {\it et al.}\cite{ref-eldredge_optimal_2016} \revision{and to the problem addressed by Sekatski {\it et al.},\cite{sekatski_optimal_2020} since one can recover the qubit nature of this example by restricting oneself to the span of the extremal eigenstates of the generators for each probe.}
Appendix~\ref{app:more-commuting} explores a zoo of variations on these sorts of measurements.

\subsection{Noncommuting generators}
\label{sec:noncommuting-generators}

Turn now to noncommuting generators.  As a simple example, consider the
Hamiltonian for a single qubit:
\begin{align}\label{eq:BlochH}
H(\theta^1,\theta^2,\theta^3)=\frac{1}{2}\big(\theta^1\sigma^x+\theta^2\sigma^y+\theta^3\sigma^z\big)
=\frac12\bvec\theta\cdot\bvec\sigma\,.
\end{align}
Here introduce, by necessity, a bastard inner product that recognizes the
natural Euclidean geometry of the Bloch sphere.  The Euclidean geometry runs
rough-shod over the distinction between upper and lower indices; using
dot notation for this inner product sidesteps ugly sums over indices that
are both upper or both lower.

The generator for a vector $\bvec b=b^j\partial_j$,
\begin{align}\label{eq:BlochY}
\Gen=\bvec b H(\theta^1,\theta^2,\theta^3)=\frac12\big(b^1\sigma^x+b^2\sigma^y+b^3\sigma^z\big)=\frac12\bvec b\cdot\bvec\sigma\,,
\end{align}
gives \QCRB\ seminorm
\begin{align}
  \Vert\bvec{b}H(\theta^1,\theta^2,\theta^3)\Vert_s
  =\sqrt{(b^1)^2+(b^2)^2+(b^3)^2}
  =\sqrt{\bvec b\cdot\bvec b}
  =\Vert\bvec{b}\Vert_2\,,
\end{align}
with $\Vert\bvec{b}\Vert_2$ being the Euclidean length of $\bvec b$.  The \QCRB\
unit circle is the Euclidean unit sphere.

Now estimate linear combination $q=q_j\theta^j$.  Tangent to the \QCRB\ sphere
the unit plane of $q$ must be; scaling $q$ appropriately, this means that
$q_j=b^j$, which also yields the desired $dq(\bvec b)=q_jb^j=\bvec b\cdot\bvec
b=1$.  The rest is standard qubitology.  Use as fiducial state an optimal state
for generator~(\ref{eq:BlochY}), say, $|\psi\rangle=\big(|\bvec
b\rangle+|{-\bvec b}\rangle\big)/\sqrt2$.  After imposition of the parameters by
Hamiltonian~(\ref{eq:BlochH}), measure in the basis $|\psi_{\bvec b}^{(\pm
i)}\rangle=\big(|\bvec b\rangle\pm i|{-\bvec b}\rangle\big)/\sqrt2$.  The
outcome probabilities,
\begin{align}
\big|\langle\psi_{\bvec b}^{(\pm i)}|e^{-iH(\bvec\theta)}|\psi\rangle\big|^2
=\frac12[1\pm\sin(\bvec{b}\cdot\bvec{\theta})]
=\frac12[1\pm\sin(q_j\theta^j)]\,,
\end{align}
depend only on the component of $\bvec\theta$ along $\bvec b$.  Realize with
satisfaction that this component---the summation convention rightly restored!---is
the property~$q$ itself, which gives the rotation angle about $\bvec b$
that is being measured.  The result?  A no-control Fisher-information matrix $F_{jk}=q_jq_k$,
whose Fisher ellipsoid consists of the two planes tangent to the unit sphere at
the tips of $\bvec b$ and $-\bvec b$.

\begin{figure}
\centering
\includegraphics[width=0.5\textwidth]{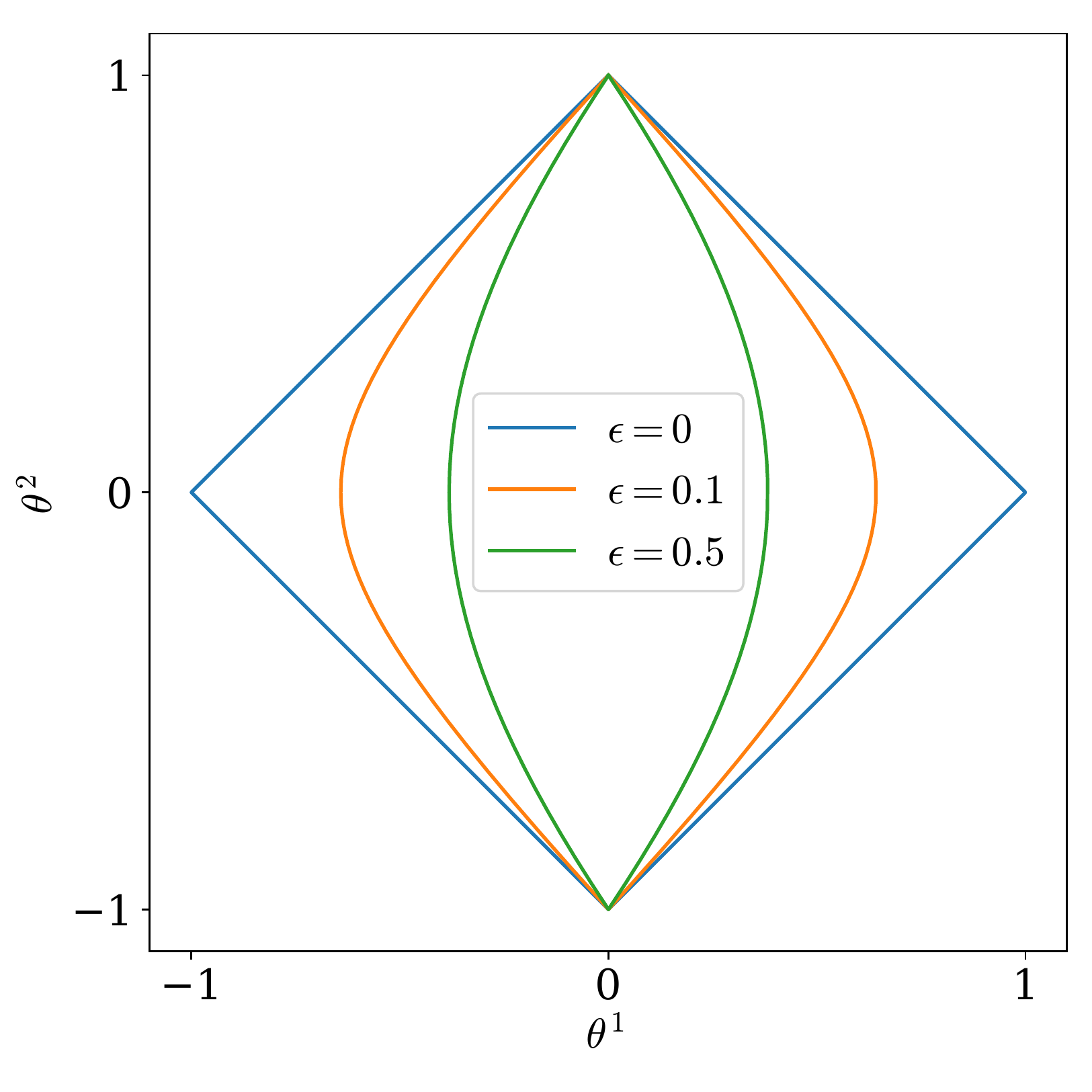}
\caption{Process norms for several generator pairs ranging from commuting
($\epsilon=0$) to increasing degrees of noncommutivity. The cross polytope
indicative of commutivity becomes rounded at two of its corners as $\epsilon$
grows.\label{fig:partially-commuting-unit-circles}}
\end{figure}

Observe more interesting behavior by varying the degree to which the generators
fail to commute, as in the two-qubit Hamiltonian
\begin{align}
  H(\theta^1,\theta^2)
  =\frac{1}{2}\big[\theta^1\big(\sigma^z_1+\sqrt{2\epsilon}\,\sigma^x_2\big)+\theta^2\sigma^z_2\big]\,.
\end{align}
The generator for vector $\bvec b=b^j\partial_j$, $\Gen=\bvec b
H(\tilde\theta)=\frac12\big[b^1\big(\sigma^z_1+\sqrt{2\epsilon}\,\sigma^x_2\big)+b^2\sigma^z_2\big]$,
has seminorm
\begin{align}
  \Vert\bvec{b}H(\theta^1,\theta^2)\Vert_s&=|b^1|
  +\sqrt{(b^2)^2+2\epsilon(b^1)^2}\,.
\end{align}
Figure~\ref{fig:partially-commuting-unit-circles} illustrates how the \QCRB\
unit circle changes as the process generators become increasingly noncommuting.
The smooth curves of the unit circle are serviced by optimal measurements like
those just encountered for the three Pauli operators; the corners present
opportunities for measurements like those encountered for commuting generators
in Sec.~\ref{sec:commuting-generators}.

To assess those opportunities, notice that vectors on the upper ($b^2\ge0$) part of
the \QCRB\ unit circle take the form
\begin{align}\label{eq:binteresting}
  \bvec{b}&=b^1\partial_1+\sqrt{1-2|b^1|+(1-2\epsilon)(b^1)^2}\,\partial_2\,.
\end{align}
The generators associated with the vectors are
\begin{align}
\begin{split}
  Y
  &=
  \frac{1}{2}\big[b^1\sigma_1^z
  +(1-\vert b_1\vert)\,\hat{\bvec{n}}\cdot\bvec{\sigma}_2\big]\,,
  \\
  \hat{\bvec{n}}
  &=
  \frac{\sqrt{1-2|b^1|+(1-2\epsilon)(b^1)^2}\,\hat{\bvec{z}}+b^1\sqrt{2\epsilon}\,\hat{\bvec{x}}}{1-\vert b^1\vert}\,.
\end{split}
\end{align}
The extremal eigenvalues of $Y$, $\pm\frac12$, correspond to eigenvectors
$|{\pm}\sgn(b^1)\hat{\bvec{z}}\rangle\otimes|{\pm}\hat{\bvec{n}}\rangle$.  The corresponding
optimal measurement is a hyperface measurement, like those in Sec.~\ref{sec:hyperface}, except
that on the second qubit the $z$ direction is replaced by $\hat{\bvec{n}}$.

Focus now on the upper cusp of the unit circle of $\Vert\bvec{b}\Vert_{\mathcal{U}_\coords}$.
Consider a scalar $q=q_1\theta^1+q_2\theta^2$, where $q_1<q_2=1$.  The unit surface of $q$
touches the \QCRB\ unit circle at the upper cusp, $\bmin=\partial_2$.  Near the cusp,
regardless of the value of $\epsilon$, the \QCRB\ unit circle looks like the square that
applies for $\epsilon=0$; for the vectors of Eq.~(\ref{eq:binteresting}), as $|b^1|\to0$,
$\hat{\bvec{n}}\to\hat{\bvec{z}}$, and the measurements are the two hyperface measurements,
for $b^1>0$ and $b^1<0$, considered in Sec.~\ref{sec:hyperface}.  Matching the tangent made by
$q$ is then carried out just as it was in Sec.~\ref{sec:kiss}.

\section{Conclusion}
\label{sec:conclusion}

Laid to rest is the question of ultimate, achievable precision in the estimation of scalar properties of arbitrary quantum channels.
Tempted to stray from the straight, but narrow path by superficial similarities to single-parameter estimation, we stayed the course
by keeping eyes fixed on the distinction between the differential forms defining our problem and the tangent vectors defining
single-parameter problems.  Yet unwise it would have been to disregard completely the voice of those who have trod the single-parameter
road, for from their stores of knowledge came forth the process norm on the tangent space.  By examining the relation
between this process norm and the differential form of the scalar property of interest, all becomes clear, and maximally precise
scalar estimation strategies emerge, beautiful to behold, constructed from the optimal single-parameter strategies known from old.

In light of these investigations of parameter estimation, as was said over two thousand years ago,
so still it must be said, ``Let no one ignorant of geometry enter here.''

\revision{
\acknowledgments
Both authors thanks the University of New Mexico's Center for Quantum Information and Control for providing a stimulating intellectual environment.  JAG was supported in part by funding from the Canada First Research Excellence Fund and from NSERC.
}

\appendix

\section{Estimator bias}
\label{app:est-bias}

Worthwhile it is to consider sensing small deviations away from a true value that
is itself close to the fiducial operating point ($\coords=0$).  Typical this
situation is, and it is the situation considered in this paper.

In this situation, calculate the Fisher information at the fiducial point,
instead of at the (unknown) true point, \ie,
\begin{align}
F_{jk}=\int dx\,p(x|\coords=0)
\left.\frac{\partial\ln p(x|\coords)}{\partial\theta^j}\right|_{\coords=0}
\left.\frac{\partial\ln p(x|\coords)}{\partial\theta^k}\right|_{\coords=0};
\end{align}
likewise, the Jacobian of the mean estimates should be calculated at the fiducial point,
\begin{align}
{J^j}_k=\left.\frac{\partial\langle\hat\theta^j\rangle_\coords}{\partial\theta^k}\right|_{\coords=0}\,.
\end{align}
Not knowing the true parameter values, we are commanded to do things this way for
the tensor formalism to make sense.

Approximate can we also the means of the estimators by expanding about the fiducial point:
\begin{align}
\langle\hat\theta^j\rangle_\coords
=\langle\hat\theta^j\rangle_{\coords=0}
+\left.\frac{\partial\langle\hat\theta^j\rangle_\coords}{\partial\theta_k}\right|_{\coords=0}\theta^k
=\langle\hat\theta^j\rangle_{\coords=0}+{J^j}_k\theta^k\,.
\end{align}
If $\hat\theta^j(x)$ is a biased estimator, an associated estimator $\hat{\bar\theta}^j(x)$ can be defined by
\begin{align}
\hat{\bar\theta}^j(x)={(J^{-1})^j}_k\big[\hat\theta^k(x)-\langle\hat\theta^k\rangle_{\coords=0}\big]\,,
\end{align}
and the new estimator is unbiased,
\begin{align}
\langle\hat{\bar\theta}^j\rangle_\coords=\theta^j\,.
\end{align}
The offset removes bias at the fiducial point; the inverse of the Jacobian
removes scaling and mixing that introduce bias away from the fiducial point.
\revision{This removal of bias in the neighborhood of the fiducial point has been called a locally unbiased estimator.\cite{suzuki_nuisance_2019,suzuki_quantum_2020,Ragy2016a}
Able to remove bias locally,} we can and should always do it and thus use the multiparameter CCRB for
unbiased estimators; it then is a matter of indifference whether we use the
error-correlation matrix or the covariance matrix to state the \hbox{CCRB}.

\section{Process norm}
\label{app:proc-norm}

Defined in Eq.~(\ref{eq:process-norm}) is the process norm.  Appreciate first
that this is a norm.  From Eq.~(\ref{eq:process-norm}), discern
that the associated unit ball is the intersection of the unit covariance ellipsoids
of all possible Fisher informations.  The length that our potential norm assigns
to any vector is the smallest positive scaling of the unit ball that contains
the vector.  A norm must assign a finite, nonnegative value to every vector;
satisfy the triangle inequality
($\Vert\bvec{v}+\bvec{w}\Vert\leq\Vert\bvec{b}\Vert+\Vert\bvec{w}\Vert$); be
absolutely scalable ($\Vert\lambda\bvec{v}\Vert=|\lambda|\Vert\bvec{v}\Vert$);
and be nondegenerate ($\Vert\bvec{v}\Vert=0\Rightarrow\bvec{v}=\bvec{0}$).  The
first three properties correspond to the unit ball being an absolutely convex
absorbing set, and the nondegeneracy corresponds to the unit ball being bounded.

A norm we have because the unit ball is absolutely convex, being an intersection
of ellipsoids, which are absolutely convex; absorbing, not assigning infinite length
to any vector $\bvec{b}$, since that would correspond to infinite estimation precision;
and bounded, not assigning zero length to any vector, since we assume that deviations
in all parameters are detectable (\ie, there are no physically meaningless
parameters).

\section{Parity measurements and $i$cat states}
\label{app:comm-gen-details}

The $i$cat states $|\psi_\tz^{(\pm i)}\rangle$ of Eq.~(\ref{eq:icat}) are eigenstates of
$(\sigma^y)^{\otimes N}$ when $N$ is odd and
$(\sigma^y)^{\otimes N-1}\otimes\sigma^x$ when $N$ is even.  Understand why by
recalling $\sigma^y|z_j\rangle=iz_j|{-}z_j\rangle$ and
$\sigma^x|z_j\rangle=|{-}z_j\rangle$.  For $N$ odd,
\begin{align}
\begin{split}
  (\sigma^y)^{\otimes N}|\psi_\tz^{(\pm i)}\rangle
  &=
  \frac{i^N}{\sqrt{2}}(z_1\cdots z_N)
  \big(|\minustz\rangle\pm(-1)^Ni|\tz\rangle\big)
  \\
  &=\mp(-1)^{(N+1)/2}(z_1\cdots z_N)|\psi_\tz^{(\pm i)}\rangle\,,
\end{split}
\end{align}
and for $N$ even,
\begin{align}
\begin{split}
  (\sigma^y)^{\otimes N-1}\otimes\sigma^x|\psi_\tz^{(\pm i)}\rangle
  &=
  \frac{i^{N-1}}{\sqrt{2}}(z_1\cdots z_{N-1})
  \big(|\minustz\rangle\pm(-1)^{N-1}i|\tz\rangle\big)
  \\
  &=\mp(-1)^{N/2}(z_1\cdots z_{N-1})|\psi_\tz^{(\pm i)}\rangle\,.
\end{split}
\end{align}
Measuring these generalized parity operators realizes an optimal measurement for
the protocols in Sec.~\ref{sec:hyperface}.

\section{Probabilistic protocols}
\label{app:convex-fisher}

Probabilistic protocols we invoke in Sec.~\ref{sec:quant-scalar-est} to argue that measurements saturating the one-from-many inequality always exist for $\bmin$.
Understand now the precise nature of these probabilistic protocols and the means by which they achieve our aim.

Given two different measurement protocols, each dictating the preparation of a particular initial state and the measurement of a particular POVM, one can combine the two by deciding to choose randomly which protocol to follow before making use of the channel of interest.
The bounds in this paper are derived allowing for the possibility of entangled ancillas.
Since random choice between different state preparation and measurement can be effected by a deterministic protocol using entangled ancillas, a probabilistic protocol along these lines is allowed within the quantum framework of Sec.~\ref{sec:QCRB}.
Entangled protocols find a place in the examples of App.~\ref{app:commutingzoo}.

The $n$th deterministic protocol has fiducial state $\rho_n$ and measures POVM $\{E_{x_n}\}$, labeled by outcomes $x_n$; the outcomes have probability $p(x_n|n,\coords)=\tr(\rho_{n,\coords}E_{x_n})$, where $\rho_{n,\coords}=\mathcal{E}_\coords(\rho_n)$ is the output of the quantum process.  The probabilistic protocol has all the outcomes of all the deterministic protocols; the probability of outcome $x_n$ is
\begin{align}
p(x_n|\coords)=p(x_n|n,\coords)p(n)\,,
\end{align}
where $p(n)$ is the probability to choose the $n$th deterministic protocol.  Now easy it is to see that the Fisher information for the probabilistic protocol is the convex combination of the Fisher informations for the deterministic protocols:
\begin{align}
\begin{split}
  F_{jk}
  &=
  \sum_n\int dx_n\,\frac{1}{p(x_n|\coords)}
  \partial_jp(x_n|\coords)
  \,\partial_kp(x_n|\coords)
  \\
  &=
  \sum_n p(n)\int dx_n\,\frac{1}{p(x_n|n,\coords)}
  \partial_jp(x_n|n,\coords)
  \,\partial_kp(x_n|n,\coords)
  \\
  &=\sum_n p(n)F_{jk}(n)\,.
\end{split}
\end{align}

Return now to the problem of constructing an optimal probabilistic protocol at a corner $\bmin$ of the \QCRB\ surface.
Consider all the $\bvec{b}$ near to $\bmin$ that have the same \QCRB\ norm.
In a small enough neighborhood, this looks like the boundary of a convex cone, called the tangent cone.
Identifying tangent planes to this cone with forms results in the construction of the dual cone.
We show that all the tangent planes to the tip of the tangent cone---that is, all forms in the dual cone---can be expressed as convex combinations of tangent planes to smooth points on the tangent cone.
Since there always exists a quantum protocol realizing at least one tangent plane to a point on the cone, and since smooth points only have one tangent plane, this implies that arbitrary tangent planes to the tip of the cone can be realized through probabilistic combinations of quantum protocols that are known to~exist.

We first eliminate irrelevant parameters so the base of the restricted tangent cone is bounded.
If the surface of constant \QCRB\ norm is flat in certain directions at $\bmin$ (for example, if it is a sphere) the tangent cone extends infinitely in that direction.
The level surfaces of $dq$ coincide exactly with the tangent cone in those directions, as does the covariance of any \QCRB-saturating measurement protocol, so we can safely ignore those directions and restrict to the remaining cone, whose base is a bounded convex set just like the unit ball of our norm.
A smooth point on the boundary of this set corresponds to a ray of smooth points on the boundary of the tangent cone.

We now argue that the set of extremal tangent planes to this restricted tangent cone is equivalent to the set of tangent planes to its base.
Extremal tangent planes are rotated out as far away as possible from being flat at the tip of the cone, so they are entirely determined by the lower-dimensional tangent plane they make with the base of the cone.
Combine this with the observation that a lower-dimensional tangent plane to a smooth point on the base corresponds to a tangent plane to a smooth point on the cone, since the additional degree of freedom in the cone is a ray emanating from the tip, and therefore smooth.

We use this trick of reducing the dimension to bootstrap a higher-dimensional protocol from lower-dimensional protocols.
Start by assuming we can make arbitrary tangent planes to any point on the boundary of this lower-dimensional convex set using a convex combination of tangent planes to smooth points in the neighborhood of that point.
From this it would follow that we can make arbitrary extremal tangent planes to the point of interest in our higher-dimensional convex set using convex combinations of tangent planes to smooth points.
Since the dual cone of tangent planes is convex, probabilistic protocols for making extremal tangent planes yield probabilistic protocols for making all tangent planes.
For a two-dimensional cone it is easy to see how to make arbitrary tangent planes to its base using convex combinations of tangent planes to smooth points, since the base is a one-dimensional object and both points on the boundary are smooth.
Inductively, one can then build up convex combinations of smooth tangent planes to construct arbitrary tangent planes of higher-and-higher-dimensional convex sets, ultimately arriving at a probabilistic protocol that matches the level surface of $dq$ at the point of interest.

Summarize: make arbitrary tangent planes to a point of interest on the unit ball by utilizing lower-dimensional protocols for making arbitrary tangent planes to points on the boundary of the base of the tangent cone of the point of interest.

\section{A zoo of measurements in the commuting case}
\label{app:more-commuting}

Hyperface measurements are the focus of Sec.~\ref{sec:hyperface},
because they are sufficient for constructing the optimal protocols needed in Sec.~\ref{sec:commuting-generators}.
Yet these are far from the only deterministic measurement
protocols that saturate the \QCRB.  As additional examples, consider hyperedges
of the cross-polytope, specified by a string $\tw=w_1\ldots w_N$, much like the
string for a hyperface, except that the characters can be $\tzero$ in addition
to $\tone$ and $\tminusone$.  The hyperedges so signified are
\begin{align}\label{eq:hyperedge}
\left\{\bvec b\,\left|\,w_jb^j=1
\quad\mbox{and}\quad
\sum_{j=1}^N|b^j|=1\right.\right\}\,.
\end{align}
In three dimensions---the cross-polytope is an octahedron---the six vertices
correspond to the six strings with two zeroes, the twelve edges to the twelve
strings with one zero, and the eight faces to the eight strings with no zeroes.
For example, $\tw=\tzero\,\mbox{$\tminusone$}\,\tzero$ is the vertex on the
negative $y$ axis, $\tw=\mbox{$\tminusone$}\,\tzero\,\tone$ is the edge that
connects the $-x$ axis with the positive $z$ axis, and
$\tw=\tone\,\mbox{$\tminusone$}\,\tone$ is the face in the octant defined by the
$+x$, $-y$, and $+z$ axes.  Generally, there are $2N$ vertices corresponding to
strings with $N-1$ zeroes; $2N(N-1)$ edges corresponding to strings with $N-2$
zeroes; $2^N$ faces corresponding to strings with no zeroes; and $2^K
N!/K!(N-K)!$ hyperedges of dimension $K$---these we call
$K$-hyperedges---corresponding to strings with $N-K$~zeroes.

Consider now achieving the \QCRB\ in a no-control estimation of $q=q_j\theta^j$
(recall that we assume that $|q_j|\le1$, $j=1,\ldots,n$) for a vector $\bvec b$
that lies on the unit surface of $q$ and also lies in the interior of a
$K$-hyperedge of the cross-polytope specified by string $\tw$.  The discussion
at Eq.~(\ref{eq:hyperedge}) leads to
\begin{align}\label{eq:wvecb}
\bvec b=\sum_{j=1}^N w_j|b^j|\partial_j\,,\qquad\Vert\bvec{b}\Vert_1=\sum_{j=1}^N|b^j|=1\,.
\end{align}
According to the discussion in Sec.~\ref{sec:quant-scalar-est}, the
cross-polytope must kiss the unit surface of $q$ at $\bvec b$.  Hence, coincide
with the $K$-hyperedge the unit surface of $q$ must, meaning that $q_j=w_j$ for
$w_j=\pm1$, with the other $q_j$s left arbitrary.  Summarize: the linear
combinations for which one-from-many and \QCRB\ can be simultaneously
saturated at $\bvec b$ of Eq.~(\ref{eq:wvecb})---notice that $dq(\bvec
b)=1$---are
\begin{align}\label{eq:qcommuting}
  q=\hspace{-10pt}\sum_{\{j\mid w_j=\pm1\}}\hspace{-10pt}w_j\theta^j
  +\hspace{-8pt}\sum_{\{j\mid w_j=0\}}\hspace{-8pt}q_j\theta^j
  =w+\hspace{-8pt}\sum_{\{j\mid w_j=0\}}\hspace{-8pt}q_j\theta^j\,,
\quad
\mbox{$|q_j|\le1$.}
\end{align}

\subsection{Measurements sensitive only to the parameters on a hyperedge}
\label{app:commutingw}

Specialize now to no-control measurements that are sensitive only to the parameters on a
hyperedge, i.e., $q=w$.  Construct the states necessary for hyperedge measurements by
considering the zero-including strings $\tw$.  Let $\twone$ be the string in which all
the zeroes in $\tw$ are replaced by $+1$:
\begin{align}
|\twone\rangle=\bigotimes_{\{j\mid w_j=\pm1\}}\hspace{-10pt}|w_j\rangle\bigotimes_{\{j\mid w_j=0\}}\hspace{-8pt}|{+1}\rangle\,.
\end{align}
Appreciate that in $\minustw$, all the zero entries remain zero, so those entries become $+1$ in $\minustwone$, giving
\begin{align}
|{\minustwone}\rangle=\bigotimes_{\{j\mid w_j=\pm1\}}\hspace{-10pt}|{-w}_j\rangle\bigotimes_{\{j\mid w_j=0\}}\hspace{-8pt}|{+1}\rangle\,.
\end{align}
Note carefully that the strings $\tw$ and $\minustw$ specify opposite $K$-hyperedges of the cross-polytope, whereas $\twone$ and $\minustwone$ specify hyperfaces that contain these opposite $K$-hyperedges, but also share hyperedges that are specified by the $1$s held in common by $\twone$ and $\minustwone$.

Introduce the analog of the cat and $i$cat states of Eqs.~(\ref{eq:cat}) and~(\ref{eq:icat}):
\begin{align}\label{eq:wcat}
|\psi_{\twone}^{(\pm)}\rangle=\frac{1}{\sqrt2}\big(|\twone\rangle\pm|{\minustwone}\rangle\big)\,,\\
|\psi_{\twone}^{(\pm i)}\rangle=\frac{1}{\sqrt2}\big(|\twone\rangle\pm i|{\minustwone}\rangle\big)\,.
\end{align}
Understand that in these states, unlike the cat and $i$cat states, the (irrelevant) qubits that have $w_j=0$ are in a product of $+1$ eigenstates of $\sigma^z$.  Any of these states is an optimal states for $\bvec b$ of Eq.~(\ref{eq:wvecb}); other optimal state can be constructed using any state for the irrelevant qubits, but the product of $+1$ eigenstates is convenient.

Use these new states as ingredients in the standard recipe.
Let the qubits begin in the state~$|\psi_{\twone}^{(+)}\rangle$.
Imposition of the parameters leads to the state
\begin{align}
e^{-iH(\tilde\theta)}|\psi_{\twone}^{(+)}\rangle
=\frac{1}{\sqrt2}\big(e^{-iw_j\theta^j/2}|\twone\rangle+e^{iw_j\theta^j/2}|{\minustwone}\rangle\big)
\exp\biggl(-\frac{i}{2}\hspace{-2pt}\sum_{\{j\mid w_j=0\}}\hspace{-8pt}\theta^j\biggr)\,.
\end{align}
The irrelevant qubits, in state $|{+1}\rangle$ in both parts of the superposition, contribute the final phase factor, which has no effect on measurement probabilities.  Make a measurement in the orthonormal basis consisting of $|\psi_{\twone}^{(\pm i)}\rangle$ and the product states $|\tz\rangle$, with $\tz\ne\twone,\minustwone$.  Results $\tz$ have zero probability, and the probabilities for the results corresponding to $|\psi_{\twone}^{(\pm i)}\rangle$ are
\begin{align}
p(\pm|\tilde\theta)
=\big|\langle\psi_{\twone}^{(\pm i)}|e^{-iH(\tilde\theta)}|\psi_{\twone}^{(+)}\rangle\big|^2
=\frac14\big|e^{-iw_j\theta^j/2}\mp i e^{iw_j\theta^j/2}\big|^2
=\frac12\big(1\pm\sin w_j\theta^j\big)\,,
\end{align}
leading to Fisher-information matrix
\begin{align}\label{eq:Fjkw}
F_{jk}
=\sum_{\pm}
\frac{1}{p(\pm|\tilde\theta=0)}
\left.\frac{\partial p(\pm|\tilde\theta)}{\partial\theta^j}\right|_{\tilde\theta=0}
\left.\frac{\partial p(\pm|\tilde\theta)}{\partial\theta^k}\right|_{\tilde\theta=0}
=w_jw_k
\end{align}
or, equivalently,
\begin{align}
\bvec{F}_\downarrow=dw\otimes dw=dq\otimes dq\,.
\end{align}
This estimation scenario gathers information only about the property $q=w=w_j\theta^j$.
For any vector $\bvec v$, we have
\begin{align}
F_{vv}=F_{jk}v^jv^k=(w_jv^j)^2=[dw(\bvec v)]^2\,.
\end{align}
which has the value $1$ for any vector on the unit surface of $w$ (the vector need not be confined to the portion of that surface that is the hyperedge $\tw$
of the polytope).
For a vector $\bvec b$ on the hyperedge specified by $\tw$, as in Eq.~(\ref{eq:wvecb}),
\begin{align}
1=dw(\bvec b)=w_jb^j=\sum_j|b^j|=\Vert\bvec b\Vert_1=\Vert\bvec b H(\tilde\theta)\Vert_s\,,
\end{align}
so the measurement satisfies the unified kissing condition~(\ref{eq:kissing-condition1}),
\begin{align}
\bvec{F}_\downarrow(\bvec{b},\bvec{v})
  =
\Vert\bvec b H(\tilde\theta)\Vert_s^2\,dq(\bvec{v})\quad\mbox{for all $\bvec{v}$}\,,
\end{align}
and is an optimal no-control measurement of the parameter $q=w$, achieving both the one-from-many bound
and the \QCRB.

\subsection{A zoo of measurements}
\label{app:commutingzoo}

Return now to a property~$q$ of the general form~(\ref{eq:qcommuting}), and visit a zoo
of varied optimal measurements that can be used for estimating $q$.

Specifying the fiducial state requires an ancillary qubit, which can be thought of
as the zeroth qubit---let it appear on the far left of tensor products---and
which does not participate in the parameter-dependent interaction.  Necessary
will it be to make one of the primary qubits special, as the primary qubit that
is entangled with the ancillary qubit, and that special qubit might as well be
the first.

Choose as fiducial state
\begin{align}\label{eq:fiducialancilla1}
|\psi\rangle=\sum_{\tz}|z_1\rangle\otimes c_\tz|\psi_\tz^{(+)}\rangle
=\sum_\tz|z_1\rangle\otimes c_\tz\frac{1}{\sqrt2}\big(|\tz\rangle+|{\minustz}\rangle\big)\,;
\end{align}
Assume the amplitudes factor as
\begin{align}
c_\tz=c_{1,z_1}c_{z_2\ldots z_N}\,;
\end{align}
squared, they are a probability distribution $p_\tz=|c_\tz|^2=p_{1,z_1}p_{z_2\ldots z_N}$.
Unpack the notation to reveal what $|\psi\rangle$ is:
\begin{align}\label{eq:fiducialancilla2}
|\psi\rangle
=\sum_{z_1}c_{1,z_1}|z_1\rangle\otimes
\sum_{z_2,\ldots,z_N}c_{z_2\ldots z_N}\frac{1}{\sqrt2}\big(|z_1,z_2,\ldots,z_N\rangle+|{-z_1,-z_2,\ldots,-z_N}\rangle\big)\,.
\end{align}
This fiducial state could be created in the following way: start the primary qubits in the $z_1=+1$ state on the right of Eq.~(\ref{eq:fiducialancilla2}), start the ancilla in the state $\sum_{z_1}c_{1,z_1}|z_1\rangle$, and run a controlled-NOT from the ancilla to the first primary qubit.

If only one $c_{1,z_1}=1$ is nonzero, the ancillary qubit is not entangled with the primary qubits, and the state of the primary qubits is a superposition of
cat states, each corresponding to opposite faces of the cross-polytope.
If all the $c_\tz=1/\sqrt{2^N}$ are equal, the ancillary qubit is not entangled with the primary qubits, and $|\psi\rangle$ reduces to
\begin{align}
|\psi\rangle=\frac{1}{\sqrt2}\big(|{+1}\rangle+|{-1}\rangle\big)\otimes\frac{1}{\sqrt{2^N}}\sum_\tz|\tz\rangle\,.
\end{align}
The sum over equal linear combination of all the basis states of the primary qubits is a product of $+1$ $\sigma^x$ eigenstates, so the entire state is a product of $+1$ $\sigma^x$ eigenstates for the ancillary qubit and the primary qubits.

Imposition of the parameters leads to
\begin{align}
\begin{split}
|\psi_{\tilde\theta}\rangle
&=e^{-iH(\tilde\theta)}|\psi\rangle\\
&=\sum_\tz|z_1\rangle\otimes\frac{1}{\sqrt2}c_{\tz}\big(e^{-iz_j\theta^j/2}|\tz\rangle+e^{iz_j\theta^j/2}|{\minustz}\rangle\big)\\
&=\frac12\sum_\tz|z_1\rangle\otimes c_\tz\Big(
\big(e^{-iz_j\theta^j/2}-ie^{iz_j\theta^j/2}\big)|\psi_\tz^{(+i)}\rangle
+\big(e^{-iz_j\theta^j/2}+ie^{iz_j\theta^j/2}\big)|\psi_\tz^{(-i)}\rangle
\Big)\,.
\end{split}
\end{align}
Measure in the orthonormal basis consisting of states $|z_1\rangle\otimes|\psi_\tz^{(\pm i)}\rangle$.
The outcome probabilities,
\begin{align}
p(\pm\tz|\tilde\theta)
=\Big|\big(\langle z_1|\otimes\langle\psi_\tz^{(\pm i)}|\big)|\psi_{\tilde\theta}\rangle\Big|^2
=\frac14 p_\tz\big|e^{-iz_j\theta^j/2}\mp i e^{iz_j\theta^j/2}\big|^2
=\frac12 p_\tz\big(1\pm\sin z_j\theta^j\big)\,,
\end{align}
give rise to Fisher-information matrix
\begin{align}\label{eq:Fjkconvexcombination}
F_{jk}
=\sum_{\pm\tz}
\frac{1}{p(\pm\tz|\tilde\theta=0)}
\left.\frac{\partial p(\pm\tz|\tilde\theta)}{\partial\theta^j}\right|_{\tilde\theta=0}
\left.\frac{\partial p(\pm\tz|\tilde\theta)}{\partial\theta^k}\right|_{\tilde\theta=0}
=\sum_\tz p_\tz z_jz_k
=\overline{z_jz_k}\,,
\end{align}
where a bar denotes an average over $p_\tz$.
This Fisher information is a convex combination of the Fisher informations of the form~(\ref{eq:Fjkw}) for the case where the two hyperedges are opposite faces of the cross-polytope.
Because $z_j^2=1$, the \revision{Fisher-information} matrix~(\ref{eq:Fjkconvexcombination}) has 1s on the diagonal.

The same result emerges if the pure fiducial state~(\ref{eq:fiducialancilla1}) is replaced by the mixed state
\begin{align}
\rho=\sum_\tz p_\tz|z_1\rangle\langle z_1|\otimes|\psi_\tz^{(+)}\rangle\langle\psi_\tz^{(+)}|\,.
\end{align}
This works because the measurement can be regarded as first determining whether $z_1$ is $\pm1$, then identifying a subspace spanned by a particular $|\tz\rangle$ and $|{\minustz}\rangle$ and then doing a measurement in the $i$cat basis $|\psi_\tz^{(\pm i)}\rangle$ within that subspace;
coherence between these possibilities matters not.

Any vector $\bvec v$ has Fisher information
\begin{align}
F_{vv}=F_{jk}v^jv^k=\sum_\tz p_\tz(z_jv^j)^2=\sum_\tz p_\tz[dz(\bvec v)]^2\,.
\end{align}
If $\bvec b$ points to a vertex on the $j$ axis of the cross-polytope, \ie, $\bvec b=\pm\partial_j$, then $dz(\bvec b)=\pm z_j$ and $F_{bb}=F_{jj}=1$.
The Fisher ellipsoid circumscribes the vertices of the cross-polytope.

Specialize now to the case where the amplitudes and probabilities factor completely,
\begin{align}
c_\tz=\prod_{j=1}^N c_{j,z_j}=c_{1,z_1}\cdots c_{N,z_N}\,,\qquad
p_\tz=\prod_{j=1}^N p_{j,z_j}=p_{1,z_1}\cdots p_{N,z_N}\,.
\end{align}
The fiducial state~(\ref{eq:fiducialancilla2}) becomes
\begin{align}\label{eq:fiducialancilla3}
|\psi\rangle
=\frac{1}{\sqrt2}
\Bigg(
\sum_{z_1}c_{1,z_1}|z_1\rangle\otimes|z_1\rangle\bigotimes_{j=2}^N\sum_{z_j}c_{j,z_j}|z_j\rangle
+
\sum_{z_1}c_{1,z_1}|z_1\rangle\otimes|{-z_1}\rangle\bigotimes_{j=2}^N\sum_{z_j}c_{j,z_j}|{-z_j}\rangle
\Bigg)\,.
\end{align}
Only the marginals
\begin{align}
\overline{z}_j=p_{j,+1}-p_{j,-1}=a_j\,,
\end{align}
which can take on values $|a_j|\le1$, matter now, with the \revision{Fisher-information} matrix becoming
\begin{align}\label{eq:Fjka}
F_{jk}=\delta_{jk}(1-a_j^2)+a_ja_k\,.
\end{align}

Worthwhile as an example is the case $a_j=a$, $j=1,\ldots,N$.
The Fisher ellipsoid has one minor axis,
\begin{align}
\bvec v=\frac{1}{\sqrt{N[1+a^2(N-1)]}}\sum_j\partial_j\,,
\end{align}
which points directly into the all-positive $2^N$-ant;
any vector $\bvec u$ that lies in the plane $0=\sum_j u^j$ and has $\sum_j(u^j)^2=(1-a^2)^{-1}$ is a major axis.
For $0<a<1$, the Fisher ellipsoid is prolate and circumscribes the cross-polytope.
When $a=0$, the Fisher ellipsoid becomes a sphere;
when $a=+1$, it degenerates to the pair of planes $\sum_j v^j=\pm1$ and thus contains the paired all-positive and all-negative faces of the cross-polytope.

Return now to the Fisher information~(\ref{eq:Fjka}).  For specificity, consider the vertex $\tw=\tone\tzero\ldots\tzero$ (the same construction works at any vertex).
Vector $\bvec b=\partial_1$ points to this vertex.  As promised by Eq.~(\ref{eq:qcommuting}), there should be an
optimal no-control measurement of the parameter ($w=\theta^1$),
\begin{align}
q=\theta^1+\sum_{j=2}^N q_j\theta^j\,,\quad\mbox{$|q_j|\le1$.}
\end{align}
Required is that the Fisher ellipsoid be tangent to the level surface of $q$;
thus demand that the gradient of the Fisher quadratic form, $F_{jk}\theta^j\theta^k$, be proportional to the gradient of $q$ at $\theta^j={\delta^j}_1$:
\begin{align}
d\theta^1+\sum_{k=2}^N q_k d\theta^k=dq
\propto d(F_{jk}\theta^j\theta^k)
=2F_{jk}\theta^j d\theta^k=2F_{1k}d\theta^k=2\bigg(d\theta^1+a_1\sum_{k=2}^N a_k d\theta^k\bigg)\,.
\end{align}
Choose
\begin{align}\label{eq:ajwone}
a_j=
\begin{cases}
1\,,&j=1,\\
q_j\,,&j=2,\ldots,N,
\end{cases}
\end{align}
to make the proportionality, and---voil{\`a}!---find a no-control procedure for estimating $q$, achieving both the one-from-many and \QCRB\ bounds.

Generalize this no-control measurement to a $K$-hyperedge.
Let $\tw=\tone\ldots\tone\tzero\ldots\tzero$, where there are \tone s in the first $K$ positions and \tzero s in the remaining $N-K$ slots (the same construction works for any $K$-hyperedge).
A vector $\bvec b$ on the hyperedge has the form~(\ref{eq:wvecb}):
\begin{align}
\bvec b=\sum_{k=1}^K b^k\partial_k\,,\qquad \sum_{k=1}^K b^k=1\,,\quad b_k\ge0\,.
\end{align}
To be estimated is a linear combination of the form~(\ref{eq:qcommuting}):
\begin{align}\label{eq:qwK}
q=w+\sum_{k=K+1}^N q_k\theta^k
=\sum_{k=1}^K\theta^k+\sum_{k=K+1}^N q_k\theta^k\,,
\quad
\mbox{$|q_k|\le1$.}
\end{align}
Any point on the $K$-hyperedge satisfies $\sum_{k=1}^K\theta^k=1$, with $\theta^k\ge0$, for $k=1,\ldots,K$,
and $\theta^k=0$, for $k=K+1,\ldots,N$.  The requirement that the Fisher ellipsoid be tangent to the level
surface of $q$ is again that at any point on the $K$-hyperedge, the gradient of the Fisher quadratic form
$F_{jk}\theta^j\theta^k$ be proportional to the gradient of $q$:
\begin{align}
\begin{split}
\sum_{k=1}^K d\theta^k\hspace{2pt}+&\sum_{k=K+1}^N q_k d\theta^k=dq\\
&\propto2F_{jk}\theta^jd\theta^k
=2\sum_{k=1}^K d\theta^k\bigg(\theta^k(1-a_k^2)+a_k\sum_{j=1}^K a_j\theta^j\bigg)
+2\hspace{-5pt}\sum_{k=K+1}^N\hspace{-5pt}d\theta^k a_k\sum_{j=1}^K a_j\theta^j\,.
\end{split}
\end{align}
Make the proportionality true by choosing
\begin{align}\label{eq:ajw}
a_j=
\begin{cases}
1\,,&j=1,\ldots,K,\\
q_j\,,&j=K+1,\ldots,N.
\end{cases}
\end{align}
Thus generalized is Eq.~(\ref{eq:ajwone}) to a no-control procedure for estimating property $q$ of Eq.~(\ref{eq:qwK}),
achieving both the one-from-many and \QCRB\ bounds.  Cylindrical is the Fisher ellipsoid for the measurement given by Eq.~(\ref{eq:ajw}): it contains the $K$-hyperedges $\tw$ and $\minustw$ and runs off to infinity along the planes defined by those hyperedges; the cross-section of the cylinder is an ellipsoid.

The choice~(\ref{eq:ajw}) is similar, yet different from the measurement formulated in App.~\ref{app:commutingw}.  The difference?
The measurement in App.~\ref{app:commutingw} uses a fiducial state that makes the measurement insensitive to parameters $\theta^k$ for $k=K+1,\ldots,N$;
the measurement here adjusts the fiducial state of the previously superfluous qubits to give just the right sensitivity to those same parameters, thus
delivering a procedure for no-control estimation of $q$ in Eq.~(\ref{eq:qwK}), instead of estimation of~$w$.

\end{document}